\documentclass[fleqn,usenatbib]{mnras}

\usepackage{newtxtext,newtxmath}

\usepackage[T1]{fontenc}
\usepackage{ae,aecompl}

\usepackage{graphicx}	
\usepackage{amsmath}	
\usepackage{amssymb}	

\newcommand{\tng}{IllustrisTNG}
\newcommand{\arepo}{\textsc{Arepo}}

\newcommand{\vmax}{${\widetilde V_{{\rm max}}}$}

\title[HOD in TNG]{Revealing the galaxy-halo connection in IllustrisTNG}

\author[S. Bose et al.]{Sownak Bose$^{1}$\thanks{E-mail: sownak.bose@cfa.harvard.edu},
Daniel J. Eisenstein$^{1}$,
Lars Hernquist$^{1}$, Annalisa Pillepich$^{2}$,\newauthor Dylan Nelson$^{3}$, Federico Marinacci$^{4}$, Volker Springel$^{3}$ and Mark Vogelsberger$^{5}$ 
\\
$^{1}$Center for Astrophysics | Harvard \& Smithsonian, 60 Garden Street, Cambridge, MA 02138, USA \\
$^{2}$Max-Planck-Institut f{\"u}r Astronomie, K{\"o}nigstuhl 17, 69117 Heidelberg, Germany \\
$^{3}$Max-Planck-Institut f{\"u}r Astrophysik, Karl-Schwarzschild-Str. 1, D-85741 Garching, Germany \\
$^{4}$Department of Physics \& Astronomy, University of Bologna, via Gobetti 93/2, 40129 Bologna, Italy \\
$^{5}$Kavli Institute for Astrophysics and Space Research, Massachusetts Institute of Technology, Cambridge, MA 02139, USA \\
}

\date{}

\pubyear{2019}

\begin{document}
\label{firstpage}
\pagerange{\pageref{firstpage}--\pageref{lastpage}}
\maketitle

\begin{abstract}
We use the \tng{} (TNG) simulations to explore the galaxy-halo connection as inferred from state-of-the-art cosmological, magnetohydrodynamical simulations. With the high mass resolution and large volume achieved by combining the 100 Mpc (TNG100) and 300 Mpc (TNG300) volumes, we establish the mean occupancy of central and satellite galaxies and their dependence on the properties of the dark matter haloes hosting them. We derive best-fitting HOD parameters from TNG100 and TNG300 for target galaxy number densities of $\bar{n}_g = 0.032\,h^3$Mpc$^{-3}$ and $\bar{n}_g = 0.016\,h^3$Mpc$^{-3}$, respectively, corresponding to a minimum galaxy stellar mass of $M_\star\sim1.9\times10^9\,{\rm M}_\odot$ and $M_\star\sim3.5\times10^9\,{\rm M}_\odot$, respectively, in hosts more massive than $10^{11}\,{\rm M}_\odot$. Consistent with previous work, we find that haloes located in dense environments, with low concentrations, later formation times, and high angular momenta are richest in their satellite population. At low mass, highly-concentrated haloes and those located in overdense regions are more likely to contain a central galaxy. The degree of environmental dependence is sensitive to the definition adopted for the physical boundary of the host halo. We examine the extent to which correlations between galaxy occupancy and halo properties are independent and demonstrate that HODs predicted by halo mass and present-day concentration capture the qualitative dependence on the remaining halo properties. At fixed halo mass, concentration is a strong predictor of the stellar mass of the central galaxy, which may play a defining role in the fate of the satellite population. The radial distribution of satellite galaxies, which exhibits a universal form across a wide range of host halo mass, is described accurately by the best-fit NFW density profile of their host haloes.
\end{abstract}




\section{Introduction}

In our current theory of structure formation, gravitational instability transforms the initially linear, near homogeneous Universe into the complex network of non-linear structures we observe around us today. The process by which tiny fluctuations in the primordial density field, inferred by microwave background experiments \citep[e.g.][]{Spergel2003,Komatsu2011,Planck2016}, are amplified into the cosmic web of filaments, voids, walls and haloes has been captured with impressive accuracy and detail by a rigorous programme of numerical simulations over the last four decades \cite[e.g.][]{Efstathiou1985,Davis1985,Springel2005,Prada2012,Angulo2012,Habib2016,Potter2017}. In particular, the statistical agreement between the clustering of galaxies measured in redshift surveys and that predicted by numerical simulations has given credence to the cold dark matter model as the standard paradigm \citep[e.g.][]{Colless2001,Cole2005,Eisenstein2005,Zehavi2011}.

At first order, the clustering of haloes does not trace the total matter distribution exactly, but is instead `biased' relative to it in a manner that correlates with the mass of the halo: clustering bias increases with increasing halo mass \citep[e.g.][]{Kaiser1984,Efstathiou1988,Cole1989,Bond1991,Mo1996,Sheth1999}. As numerical simulations have allowed us to probe deeper into the non-linear regime, it has become feasible to quantify other properties of haloes -- in addition to their mass -- that may also determine this bias. The most commonly-explored properties in this context are the large-scale environment in which the halo is embedded, its formation time, concentration and spin; the dependence of clustering on these secondary quantities is broadly categorised under the umbrella term of `assembly bias' \citep[e.g.][see \citealt{Desjacques2018} for a recent review]{Gao2005,Harker2006,Wechsler2006,Bett2007,Gao2007,Jing2007,Zentner2007,Dalal2008,Mao2018}. 

Clustering bias is not merely a theoretical concept, but is also imprinted in the observed distribution of galaxies. Correlations between galaxy properties like colour, luminosity, star formation history and morphology of galaxies and their large-scale environment has long been known from redshift surveys \citep[e.g.][]{Davis1976,Dressler1980,Blanton2005}. \cite{Zehavi2005,Zehavi2011} demonstrated a strong luminosity and colour dependence of galaxy clustering in Sloan Digital Sky Survey, while an analogous correlation with age and star formation rate has been established recently by e.g. \cite{Wang2013,Hearin2013}. The combined use of photometric galaxy catalogues and weak gravitational lensing by rich clusters has indicated the existence of assembly bias as a function of concentration and assembly history \citep[e.g.][]{Miyatake2016,More2016,Montero2017,Niemiec2018}, although it has been suggested that foreground contamination from interloping galaxies introduces systematics that manifest falsely as assembly bias \citep{Busch2017}. 

On scales smaller than $\sim$10 Mpc or so, comparison between theory and observation is muddied by the complex interplay between dark matter and baryons. Hydrodynamical simulations have shown that baryonic processes alter the clustering of matter inferred from purely collisionless simulations at the level of tens of per cent and in a scale-dependent fashion \citep[e.g.][]{vanDaalen2014,vanDaalen2015,Hellwing2016,Chisari2018,Springel2018}. As galaxies are themselves biased tracers of the underlying density field, a faithful assessment of the impact of assembly bias on galactic populations requires a comprehensive model to associate them with their host haloes \citep[see][for a detailed review of this subject]{Wechsler2018}.

A common prescription of this kind, broadly referred to as semi-analytic modelling, follows merger trees to grow galaxies within dark matter haloes using systems of coupled differential equations describing gas cooling, star formation and feedback \citep[e.g.][]{Kauffmann1993,Somerville1999,Cole2000,Croton2006,Benson2012,Henriques2015}. An alternative approach involves the statistical assignment of galaxies to haloes, the most prominent of which includes abundance matching \citep[e.g.][]{Mo1999,Kravtsov2004,Tasitsiomi2004,Vale2004,Behroozi2010,Reddick2013}, empirical models \citep[e.g.][]{Conroy2009,Behroozi2013,Moster2013,Rodriguez2016,Tacchella2018,Behroozi2018} and Halo Occupation Distributions \citep[HODs,][]{Peacock2000,Benson2000,Berlind2002,Zheng2005}. Finally, hydrodynamical codes that self-consistently track the evolution of baryons and dark matter allow us to infer the association between galaxies and haloes on sub-Mpc scales; the sophistication and accuracy of these models have improved greatly over the last decade \citep[e.g.][]{Governato2009,Guedes2011,Hopkins2014,Dubois2014,Vogelsberger2014,Schaye2015,Dave2016,Pillepich2018}.

Despite the impressive improvement in the performance of hydrodynamical codes, however,the minimum galaxy stellar mass imposed by the number density of each catalogue is $\log\, [M_\star/{\rm M}_\odot] = 9.28$ in TNG100 and $\log\, [M_\star/{\rm M}_\odot] = 9.55$ their application to the domain of large-scale structure is limited by the computational cost of simulating such large volumes. Impending surveys like Euclid \citep{Laureijs2011} and the Dark Energy Spectroscopic Instrument \citep[DESI,][]{Levi2013} demand the generation of several mock simulations at Gpc scale for an accurate assessment of covariance matrices, systematics induced by cosmological parameters etc. \citep{Percival2014,OConnell2016,Klypin2018}. The spatial and temporal resolution required by hydrodynamical simulations makes this enterprise currently unfeasible. On the other hand, the HOD formalism provides a convenient (and, computationally cheap) framework to `paint' a mock galaxy population on top of large volume collisionless simulations. Cast in its original form, the HOD formalism is agnostic about any property, apart from mass, when determining the average occupancy of galaxies in haloes; assembly bias is, by construction, excluded. This first-order approximation works surprisingly well: indeed, the clustering of galaxies selected by their luminosity or colour is well reproduced by HODs defined by halo mass only \citep{Tinker2008,Tinker2009}.

In recent times, the development of `decorated' HODs \citep[e.g.][]{Hearin2016,Yuan2018} has augmented the standard HOD paradigm to incorporate scatter that may be introduced by assembly bias. Besides assembly bias, it has also been suggested that the {\it shape} of the HOD may be sensitive to physical processes associated with galaxy formation operating in host haloes, such as the strength of feedback from active galactic nuclei \citep[AGN,][]{McCullagh2017}. Apart from the demographics of galaxies, assembly bias may also play an important role in predicting differential quantities associated with galaxy formation, such as colour or star formation rate, over scales that extend well beyond the virial radius of host haloes \citep[e.g.][]{Weinmann2006,Hearin2013,Hearin2015,Lacerna2014,Kauffmann2015,Bray2016}.
 
Our goal in this paper is to exploit modern cosmological, hydrodynamical simulations to assess how assembly bias enters the HOD inferred from these simulations. In particular, we use the \tng{} suite of simulations\footnote{\href{http://www.tng-project.org/}{http://www.tng-project.org/}} to match haloes in a fully hydrodynamical simulation to their counterparts in a collisionless setup and investigate how the average HOD responds to properties like environment, concentration, formation time and spin. This exercise is similar in spirit to the recent explorations of \cite{Zehavi2018} and \cite{Artale2018} and complements them with an evaluation of the relative importance of the individual biases. At fixed halo mass, the abundance of subhaloes in dark matter-only simulations are known to vary with formation time and concentration \citep[e.g.][]{Gao2004}.

The layout of this paper is as follows. In Section~\ref{sect:numerical}, we present our numerical setup, describing the simulations and the procedures used to explore the HOD in \tng{}. Section~\ref{sect:results} presents our main findings. Finally, our conclusions are summarised in Section~\ref{sect:conclusions}.

\section{Numerical preliminaries}
\label{sect:numerical}

In this section, we briefly introduce the \tng{} magnetohydrodynamical simulations (Section~\ref{sect:sims}), which are used as the theoretical framework for exploring the Halo Occupation Distribution (HOD). We then describe the methodology used to associate galaxies and their host haloes to their counterparts in simulations containing only dark matter (Section~\ref{sect:matching}) and how the HODs are then constructed (Section~\ref{sect:construct}). Finally, we describe and quantify the list of halo properties used to assess the scatter in the HOD (Section~\ref{sect:haloprops}).

\subsection{Simulations}
\label{sect:sims}

{\it The Next Generation Illustris} (\tng{}, hereafter simply TNG) simulations \citep{Pillepich2018b,Nelson2018a,Marinacci2018,Naiman2018,Springel2018} analysed in this paper constitute an ambitious programme of cosmological, hydrodynamical simulations of galaxy formation. The simulations have been performed using the \arepo{} simulation code \citep{Springel2010}, which employs a hybrid tree/particle-mesh scheme to solve for gravitational interactions of dark matter particles, and a moving, unstructured Voronoi mesh to solve equations of hydrodynamics. A key feature of \arepo{} is that the moving mesh is adaptive in nature, resolving fluids in regions of high density with many more cells of a smaller size than in low density environments.

The TNG simulations are specified by a comprehensive galaxy formation model that has been incorporated into \arepo{}. The details of the physics model are described in detail in \cite{Weinberger2017} and \cite{Pillepich2018} and, for the sake of brevity, are not repeated here. The TNG model is the direct successor to the original \textsc{Illustris} model \citep{Vogelsberger2013,Vogelsberger2014a,Vogelsberger2014} with several updates made to the implementation of AGN feedback \citep{Weinberger2017}, the operation of galactic winds \citep{Pillepich2018}, and the incorporation of magnetic fields \citep{Pakmor2011,Pakmor2013,Pakmor2014}. This model has been shown to reproduce a wide range of properties of observed galactic populations across cosmic time \citep{Nelson2018a,Pillepich2018b,Springel2018,Naiman2018,Marinacci2018}.

In this work we make use of the TNG100 and TNG300 simulation volumes; both datasets have been made publicly available\footnote{\href{http://www.tng-project.org/data/}{http://www.tng-project.org/data/}} \citep{Nelson2018b}. The TNG100 run consists of a periodic box of length $L_{{\rm box}} = 75\,h^{-1}$Mpc $\approx$ 100 Mpc, containing 2$\times$1820$^3$ dark matter particles / gas cells, corresponding to an effective mass resolution of $9.44\times10^5\,h^{-1}\,{\rm M}_\odot$ in baryons and $5.06\times10^6\,h^{-1}\,{\rm M}_\odot$ in dark matter. The maximum physical softening length of dark matter and star particles is set to $0.5\,h^{-1}$kpc. TNG300 simulates a larger cosmological box of size $L_{{\rm box}} = 205\,h^{-1}$Mpc $\approx$ 300 Mpc with 2$\times$2500$^3$ resolution elements. The corresponding mass resolution is $7.44\times10^6\,h^{-1}\,{\rm M}_\odot$ in baryonic matter and $3.98\times10^7\,h^{-1}\,{\rm M}_\odot$ in dark matter. The maximum physical softening length of dark matter and star particles is set to $1.0\,h^{-1}$kpc.

Initial conditions for both sets of simulations have been generated at $z=127$ assuming cosmological parameters inferred by {\it Planck} \citep{Planck2016}: $\Omega_0 = 0.3089$ (total matter density), $\Omega_{\rm b} = 0.0486$ (baryon density), $\Omega_\Lambda = 0.6911$ (dark energy density), $H_0 = 67.74$ kms$^{-1}$Mpc$^{-1}$ (Hubble parameter) and $\sigma_8 = 0.8159$ (linear rms density fluctuation in a sphere of radius 8 $h^{-1}$ Mpc at $z=0$). Each of the TNG100 and TNG300 simulations have counterparts generated from the same initial phases, but evolved with dark matter only (DMO); these simulations are labelled TNG100-DMO and TNG300-DMO, respectively, and provide the properties of dark matter haloes used to vary the HOD inferred from the runs with full physics (see Section~\ref{sect:haloprops}). 

\subsection{Identifying and matching haloes}
\label{sect:matching}

Haloes are identified from the particle distribution first using a `friend-of-friends' (FOF) algorithm, which connects together dark matter particles separated by at most 20\% the mean interparticle separation to form groups \citep{Davis1985}. Particles within each FOF group that are truly gravitationally bound are then identified using the \textsc{subfind} algorithm \citep{Springel2001}, with the requirement that each `sub'halo contains at least 20 resolution elements, regardless of type. This splits a FOF group into a `main' halo and its associated subhaloes; a galaxy is then defined by the constituent stars, gas, black holes and dark matter of a (sub)halo, either central or satellite. Unless otherwise specified, a galaxy is assigned as a member of a FOF group only if it is located within $r_{200}$ of the main halo centre, which we determine in post-processing. Here, $r_{200}$ is defined as the radius within which the mean density of the halo is equal to 200 times the critical density of the universe at the given redshift. In what follows, the mass of a dark matter halo is quoted in terms of $M_{200}$ (i.e., the total mass contained within $r_{200}$); the stellar mass of a galaxy is the mass contained within a 30 kpc aperture centred on the galaxy.

We establish matches between haloes in the TNG runs with DMO and those with full physics using the procedure outlined in e.g. \cite{Lovell2018} and \cite{Bose2018}. First, we consider the 50 most-bound dark matter particles from a candidate halo in the hydrodynamical run, and search for the DMO halo in which there are at least 25 (50 per cent) of these particles. The match is then confirmed by repeating the same process, this time starting with the DMO haloes. More than 97 per cent of haloes more massive than $\sim 10^{11}\,{\rm M}_\odot$ are matched successfully using this bijective scheme. In terms of the HOD, the total number of galaxies assigned to a halo in the DMO simulation is then simply equal to the number of galaxies associated with its match in the hydrodynamical simulation.

\begin{figure}
    \centering
    \includegraphics[width=\columnwidth]{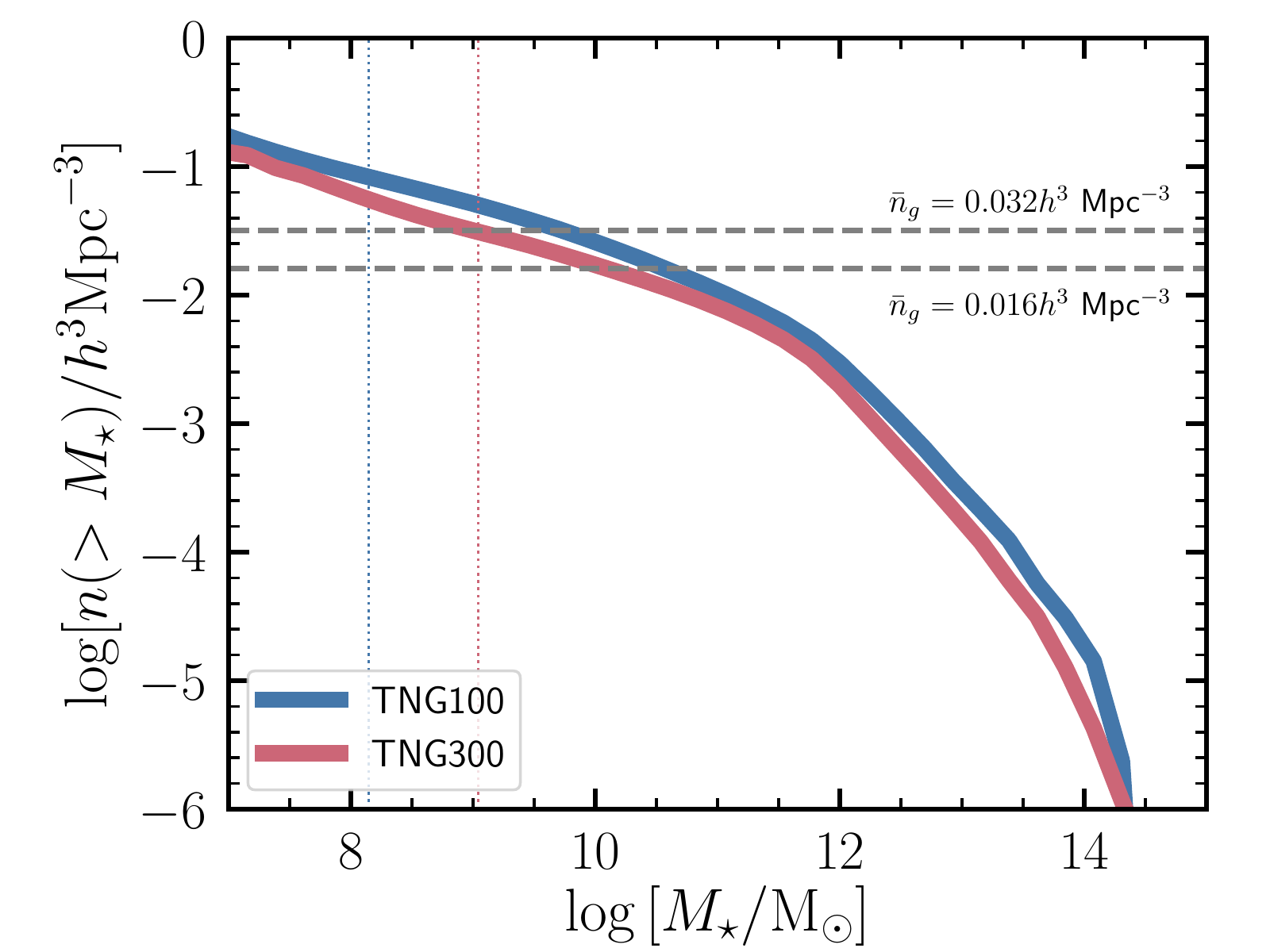}
    \caption{Cumulative number density of galaxies as a function of stellar mass in the TNG100 and TNG300 simulations. The vertical dotted lines mark the mass scale defined by 100 star particles in each simulation. The galaxy number density, $\bar{n}_g$, used to construct the HOD catalogue in TNG100 is $\bar{n}_g = 0.032\,h^3$ Mpc$^{-3}$, while for TNG300, we use $\bar{n}_g = 0.016\,h^3$ Mpc$^{-3}$; these are denoted by the horizontal dashed lines. In both instances, we are comfortably above the 100 star particle threshold.}
    \label{fig:smf}
\end{figure}

\begin{table*}
     \centering
     \begin{tabular}{c|c|c|c|c|c|c|c}
    \hline \hline
         Simulation & Number density, $\bar{n}_g$ & $\log M^{\rm  min}_\star$ & $\log M_{{\rm min}}$ & $\sigma_{\log M}$ & $\log M_{{\rm cut}}$ & $\log M_1$ & $\alpha$\\
         \hline
         (a) TNG100 & 0.032 $h^3$ Mpc$^{-3}$ & 9.28 & 11.292 & 0.165 & 11.589 & 12.483 & 1.017\\
         (b) TNG100 & 0.016 $h^3$ Mpc$^{-3}$ & 9.84 & 11.587 & 0.157 & 11.851 & 12.795 & 1.029\\
         (c) TNG300 & 0.016 $h^3$ Mpc$^{-3}$ & 9.55 & 11.601 & 0.161 & 11.778 & 12.809 & 1.018\\
         \hline
     \end{tabular}
     \caption{Best-fitting HOD parameter values for TNG100 and TNG300 at their respective target galaxy number densities (see main text for descriptions of these parameters). The third column, $\log M^{\rm min}_\star$, lists the minimum galaxy stellar mass implied by each choice of number density. The main analysis in this paper focuses on (a) and (c) only.}
     \label{tab:HOD_fit_params}
 \end{table*}

\subsection{Constructing the Halo Occupation Distribution}
\label{sect:construct}

To construct the HOD from each simulation, we first rank order galaxies from the hydrodynamical version of each simulation volume in order of decreasing stellar mass. We then select the first $N$ galaxies out of this ranked catalogue, where $N=\bar{n}_g L_{{\rm box}}^3$ and $\bar{n}_g$ is a free parameter corresponding to the target number density of galaxies we wish to construct the HOD for. For each galaxy that enters the selection, we record its type (central or satellite) and the properties of the DMO halo which its hydrodynamical host is matched to. 

To be confident of any conclusions we derive from the simulations, it is important that we limit the galaxy catalogues to entities that are well-resolved. Fig.~\ref{fig:smf} shows the (cumulative) number density of galaxies at $z=0$ in TNG100 and TNG300 as a function of stellar mass. The vertical dotted lines mark the stellar mass limit corresponding to 100 particles at the resolution limit of each simulation volume. The horizontal dotted lines, corresponding to $\bar{n}_g = 0.032\,h^3$Mpc$^{-3}$ and $\bar{n}_g = 0.016\,h^3$Mpc$^{-3}$ are, respectively, the target number densities used to construct the HOD from TNG100 and TNG300. These number densities are chosen so as to facilitate comparison with studies by \cite{Zehavi2018} and \cite{Artale2018} who adopt similar values. Correspondingly, the minimum galaxy stellar mass imposed by the number density of each catalogue is $\log\, [M^{\rm min}_\star/{\rm M}_\odot] = 9.28$ in TNG100 and $\log\, [M^{\rm min}_\star/{\rm M}_\odot] = 9.55$ in TNG300; unless otherwise stated, galaxies less massive than these limits are not considered in the remainder of our investigation. The comparatively higher mass resolution of TNG100 affords us the possibility to construct a galaxy sample with a higher number density than with TNG300; on the other hand, the larger volume of TNG300 enables us to better sample the HOD in the regime of rich clusters of galaxies ($\log\,[M_{200}/{\rm M}_\odot] \gtrsim 14.5$). Both choices of $\bar{n}_g$ result in catalogues resolved with many more than 100 star particles per galaxy. Note that the lack of overlap between the two curves presented in Fig.~\ref{fig:smf} is expected given the comparatively poorer resolution of TNG300 compared to TNG100. In particular, worse resolution results in somewhat lower stellar mass formed at fixed halo mass \citep[see][for a detailed discussion]{Pillepich2018b}. The conclusions we derive on the HOD are robust to these differences.

\begin{figure}
    \centering
    \includegraphics[width=\columnwidth]{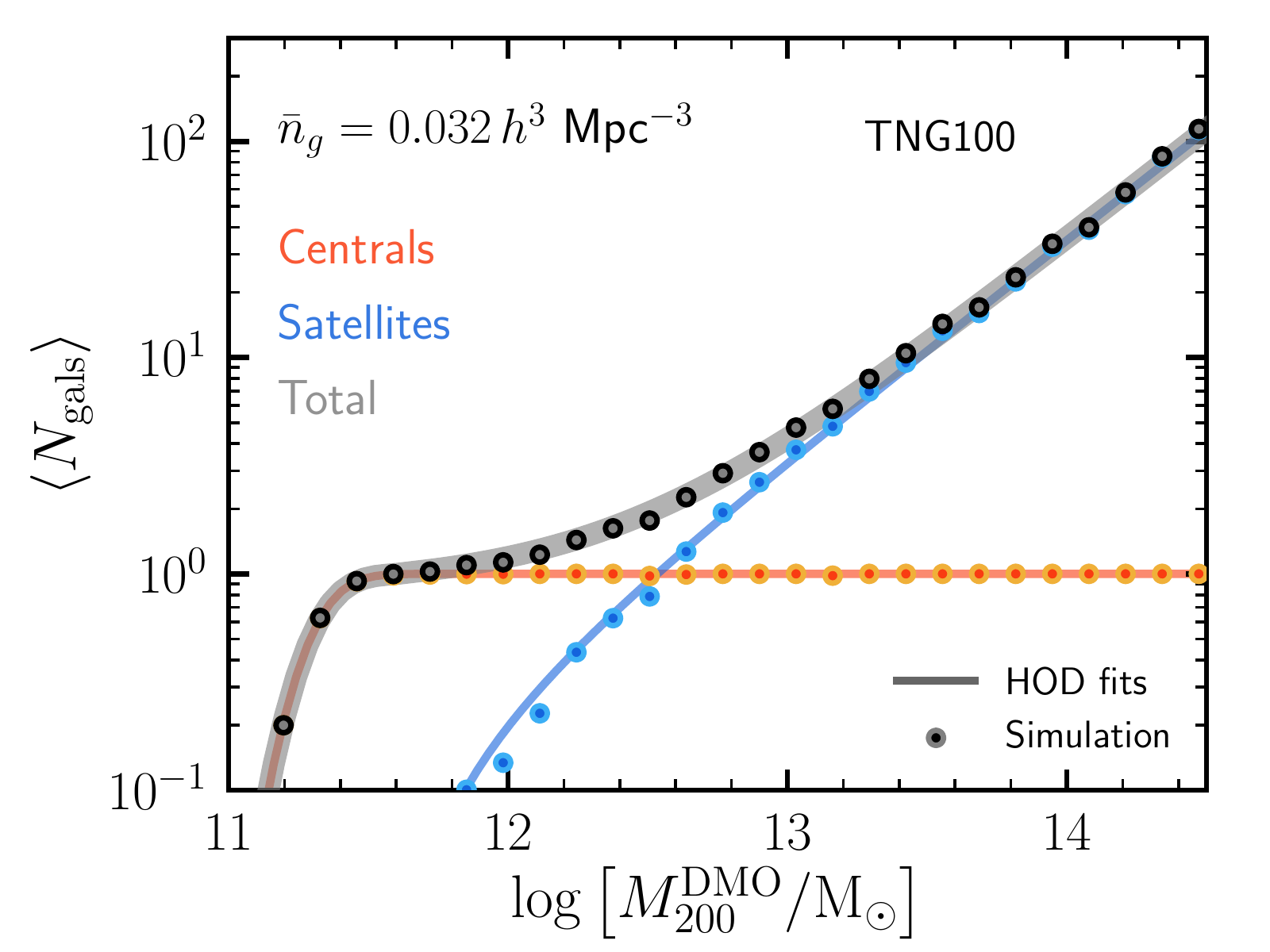} \\
    \includegraphics[width=\columnwidth]{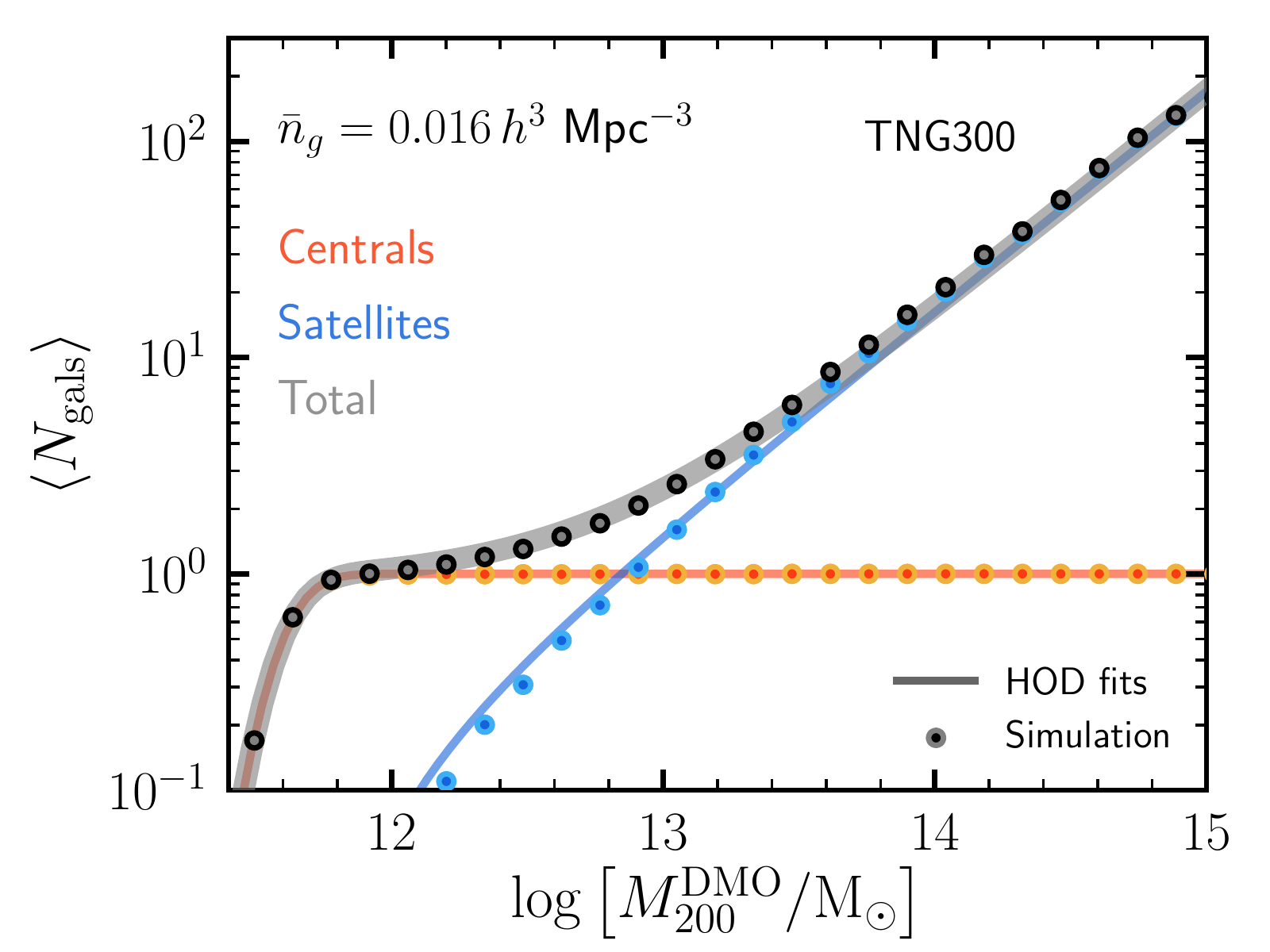}
      \caption{The mean Halo Occupation Distribution (HOD) in the TNG100 (top) and TNG300 (bottom) simulations. Each HOD is constructed at fixed number density as indicated in the top left corner of each panel. The mean occupation of central galaxies is shown in orange, satellite galaxies in blue and the total population in black. Symbols represent the mean occupancy measured in the TNG simulation volumes; the solid curves are fits assuming the \citealt{Zheng2005} model.}
    \label{fig:HOD_fits}
\end{figure}

Fig.~\ref{fig:HOD_fits} presents the HODs constructed from TNG100 (top panel) and TNG300 (bottom panel) obtained after rank-ordering galaxies and associating them with their host DMO haloes as described in the previous subsection. The HODs are constructed at target number densities of $\bar{n}_g = 0.032\,h^3$Mpc$^{-3}$ and $\bar{n}_g = 0.016\,h^3$Mpc$^{-3}$, respectively, for TNG100 and TNG300. We split the mean occupation per halo mass, $\left<N_{{\rm gals}}\right>$, into contributions from central, $\left<N_{{\rm cen}}\right>$, and satellite galaxies, $\left<N_{{\rm sat}}\right>$, with the values measured from TNG represented by the symbols. On average, every halo in TNG100-DMO more massive than $\log\, [ M_{200}^{{\rm DMO}} /{\rm M}_\odot] \sim 11.5$ hosts at least a central galaxy, while the propensity to contain at least one satellite galaxy appears at around an order of magnitude higher. At the chosen galaxy number density of $\bar{n}_g = 0.016\,h^3$Mpc$^{-3}$, only one in ten TNG300-DMO haloes, on average, contain a central galaxy. 
The mass scales of interest may be better quantified by fitting the simulation results with parameterised versions of the HOD model. The solid lines in Fig.~\ref{fig:HOD_fits} show fits to these populations assuming the 5-parameter HOD model described in \cite{Zheng2005}, in which:
\begin{equation}\label{eq:ncen}
    \left< N_{{\rm cen}} (M_h) \right> = \frac{1}{2} \left[ 1 + {\rm erf} \left( \frac{\log M_h-\log M_{{\rm min}}}{\sigma_{{\log M}}} \right) \right]\;,
\end{equation}
and
\begin{equation}\label{eq:nsat}
    \left<N_{{\rm sat}} (M_h)\right> = \left( \frac{M_h-M_{{\rm cut}}}{M_1} \right)^\alpha\;.
\end{equation}
Here, $M_h=M_{200}^{{\rm DMO}}$ is the halo mass, $M_{{\rm min}}$ is the characteristic minimum mass of haloes that host central galaxies, and $\sigma_{\log M}$ is the width of this transition. Furthermore, $M_{{\rm cut}}$ is the characteristic cut-off scale for hosting satellites, $M_1$ is a normalisation factor and $\alpha$ is the power-law slope. Eqs.~\ref{eq:ncen} \&~\ref{eq:nsat} capture the overall shape of the HOD from our simulations extremely well; the corresponding values for the 5 free parameters of this model ($M_{{\rm min}}$, $\sigma_{{\log M}}$, $M_{{\rm cut}}$, $M_1$ and $\alpha$) are listed in Table~\ref{tab:HOD_fit_params}. The middle row measures these parameters for TNG100 at the same number density as the one adopted for TNG300: while the values are quite similar, there are minor differences that may be attributed to the difference in numerical resolution between the two volumes. Throughout the rest of this paper, we will be concerned with the HOD measured at $z=0$ only. The redshift evolution of HOD fitting parameters has been studied extensively in recent work by e.g. \cite{Contreras2017} and \cite{Smith2017}.

\subsection{Statistics of the dark halo population}
\label{sect:haloprops}

The primary goal of the present investigation is to identify properties of the dark matter halo population (in addition to total mass) that are most informative of the mean (and scatter in) occupancy of galaxies within these haloes in a full hydrodynamical simulation. In other words: what properties of a halo in a DMO simulation determine if its hydrodynamical counterpart is more/less likely to host a central, or contains more/fewer satellite galaxies than the {\it average} halo at {\it that mass}.

In order to narrow down the search space for such parameters, it is useful to outline a set of criteria that these quantities should ideally satisfy. First, it is important to avoid spurious correlations in the HOD by considering only properties that have a plausible connection to galaxy occupancy. For example, a halo's epoch of formation is likely to be an informative parameter; its triaxiality, less so (at least directly). We further prioritise halo properties that are well-defined quantitatively, not very sensitive to resolution and ideally measurable at $z=0$. These conditions are especially pertinent for lower resolution, large volume simulations of large-scale structure where HOD modelling is likely to be applied, but where, for example, the construction of high resolution merger trees is not possible.

With these considerations in mind, we focus on the following set of halo properties:

\begin{figure*}
    \centering
    \includegraphics[width=0.475\textwidth]{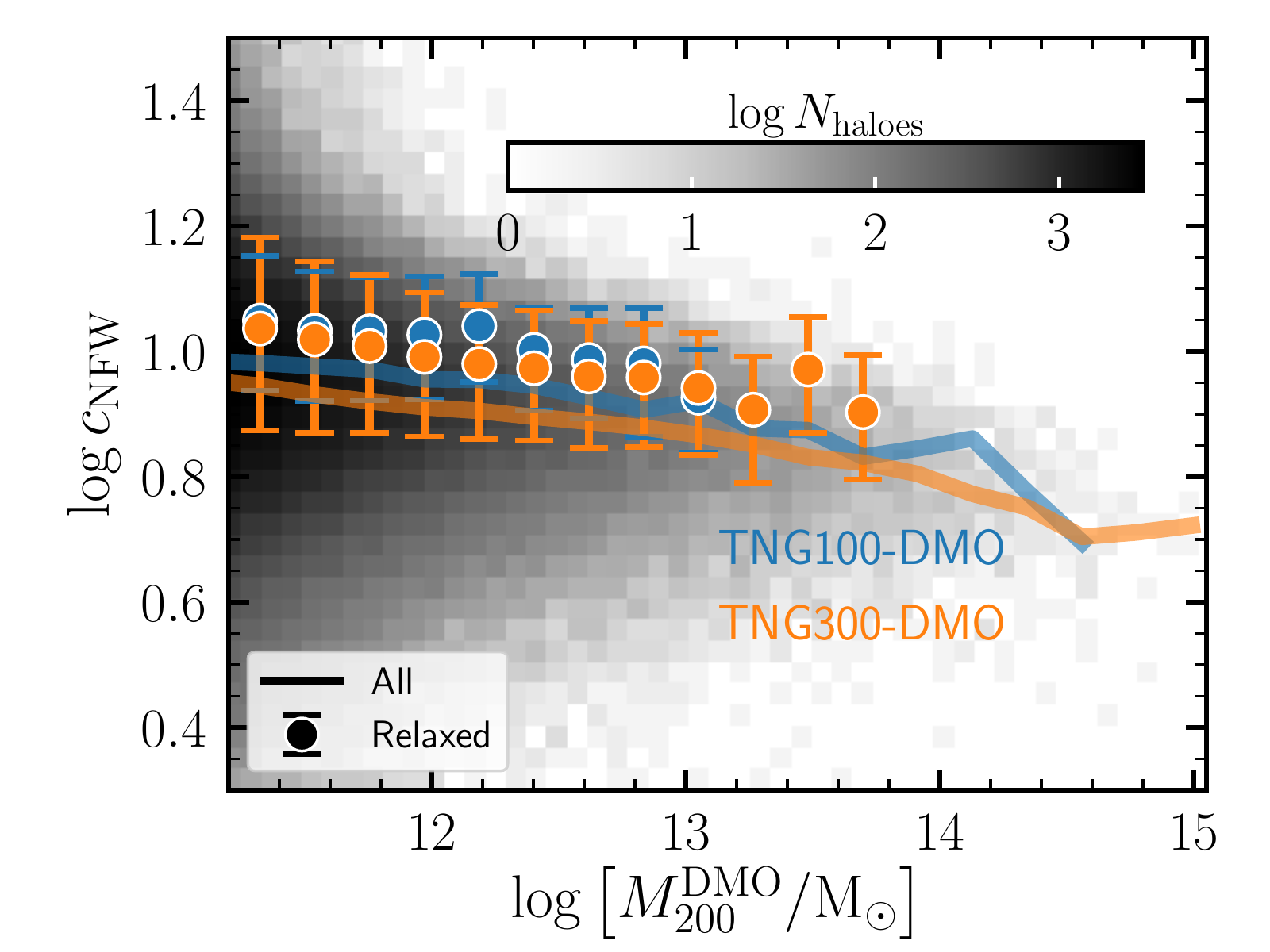}
    \includegraphics[width=0.475\textwidth]{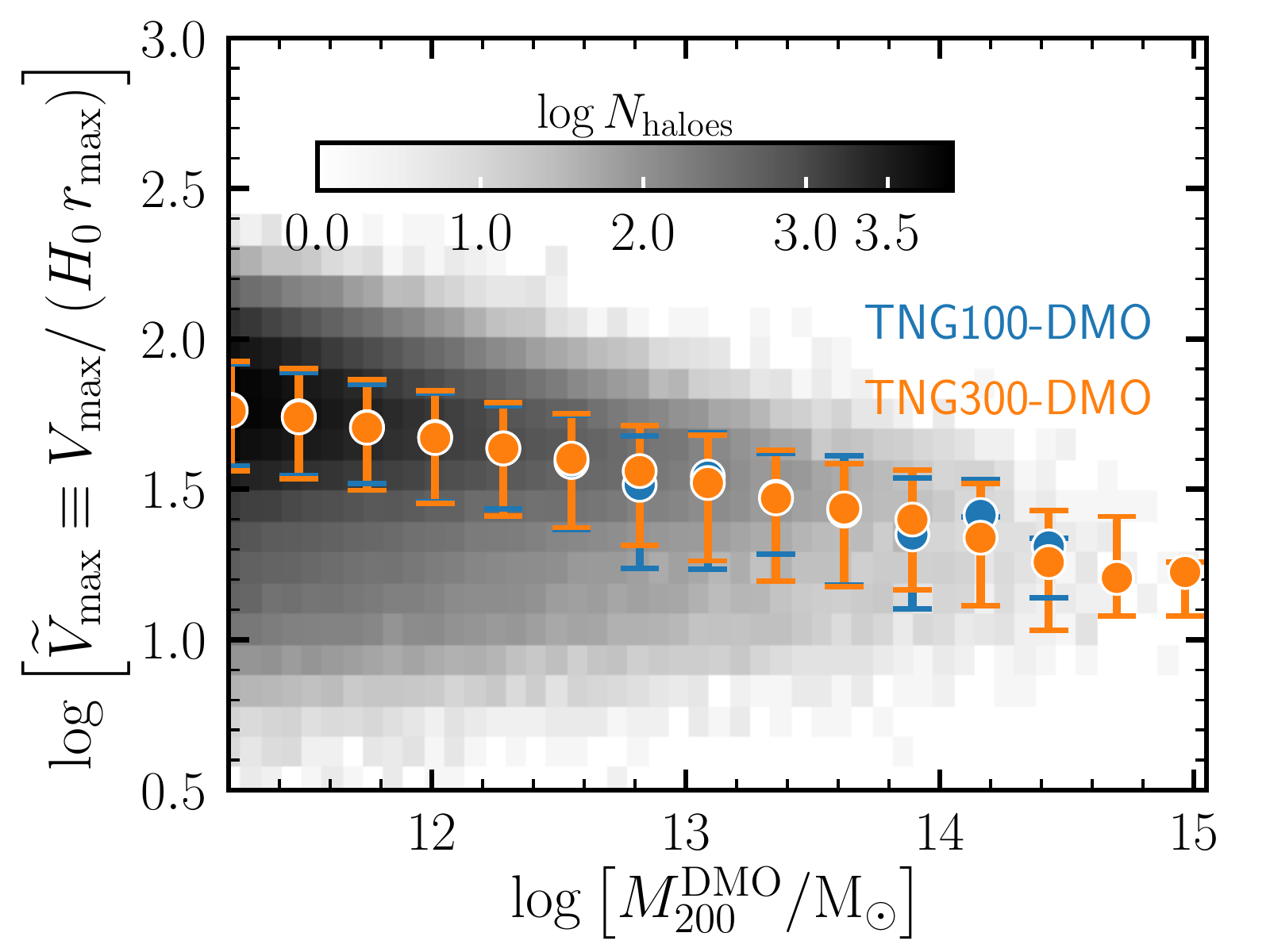} \\
    \includegraphics[width=0.475\textwidth]{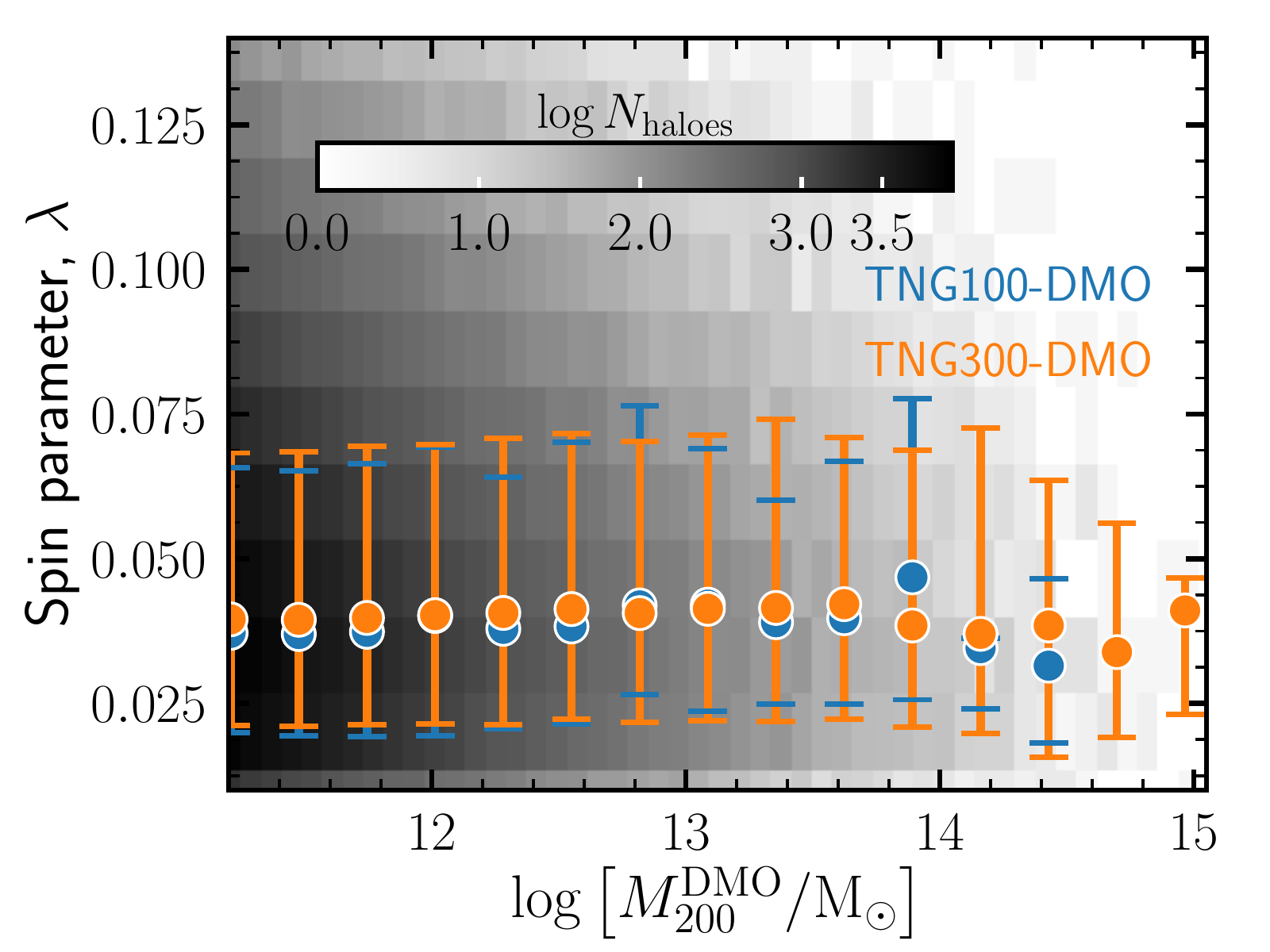}
    \includegraphics[width=0.475\textwidth]{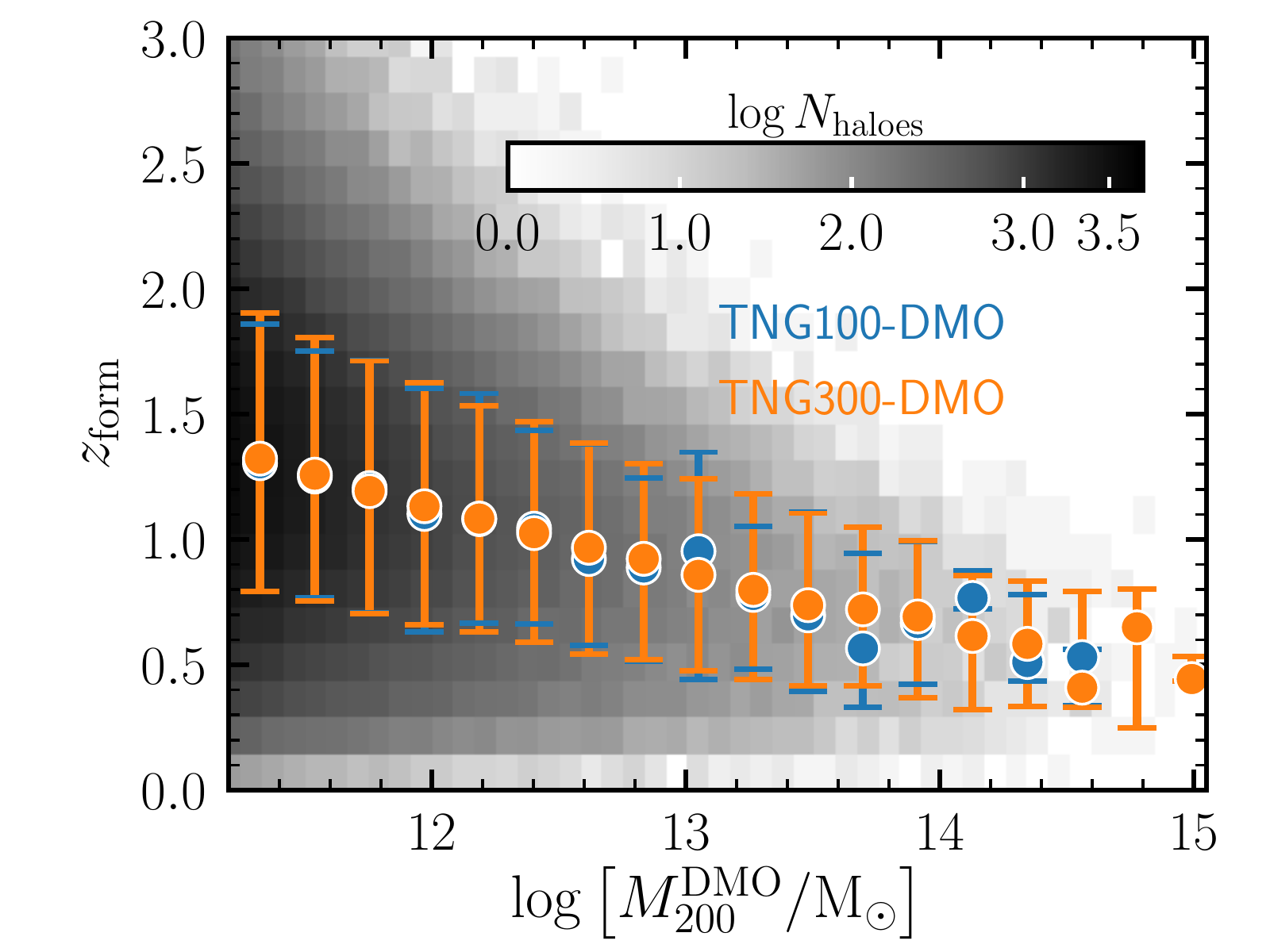}\\
    \caption{Properties of haloes extracted from the dark matter-only (DMO) versions of the TNG100 and TNG300 simulation volumes. In each panel, marker symbols show the median relation, while error bars encompass the 16$^{{\rm th}}$ and 84$^{{\rm th}}$ percentiles. All relations are shown for $z=0$. The grey histogram in the background reveals the number of haloes contained in each mass bin for the TNG300 simulation. {\it Top left panel}: the mass-concentration relation, where the halo concentration, $c_{{\rm NFW}}$, is determined by fitting NFW profiles to density profiles of individual dark matter haloes. {\it Top right panel}: Recasting the mass-concentration relation in terms of the quantity ${\widetilde V_{{\rm max}}} \equiv V_{{\rm max}}/\left(H_0\,r_{{\rm max}}\right)$ (see main text). The behaviour of this relation -- in terms of the trend and size of scatter as a function of $M_{200}^{{\rm DMO}}$ -- is qualitatively similar to the traditional mass-concentration relation. {\it Bottom left panel}: correlation between the dimensionless halo spin, $\lambda$, and halo mass. {\it Bottom right panel}: the halo formation time, $z_{{\rm form}}$, as a function of its mass, where the formation redshift is defined as the epoch by which 50\% of the halo's present day mass has been assembled.}
    \label{fig:dmo_props}
\end{figure*}

\begin{figure*}
    \centering
    \includegraphics[width=0.95\textwidth]{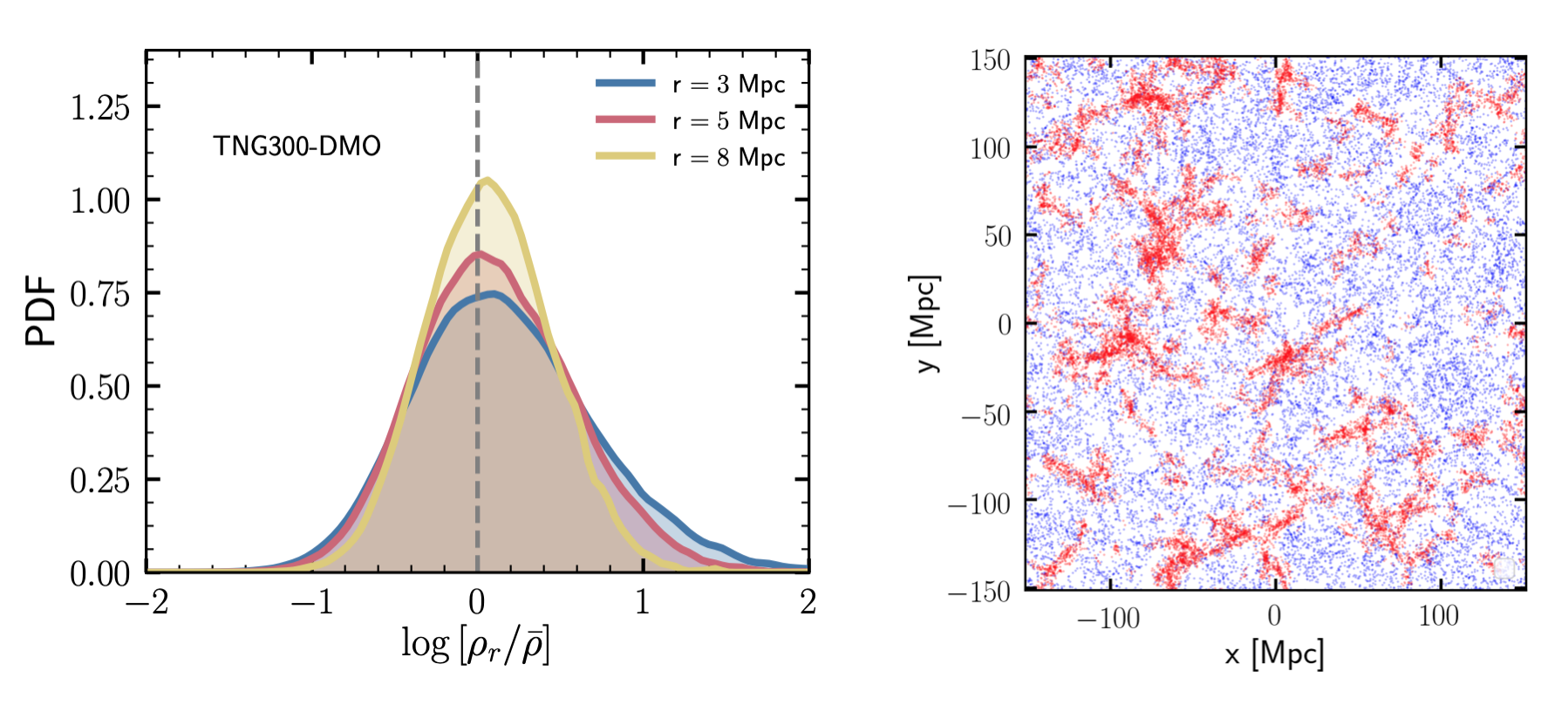}\\
    \includegraphics[width=0.95\textwidth]{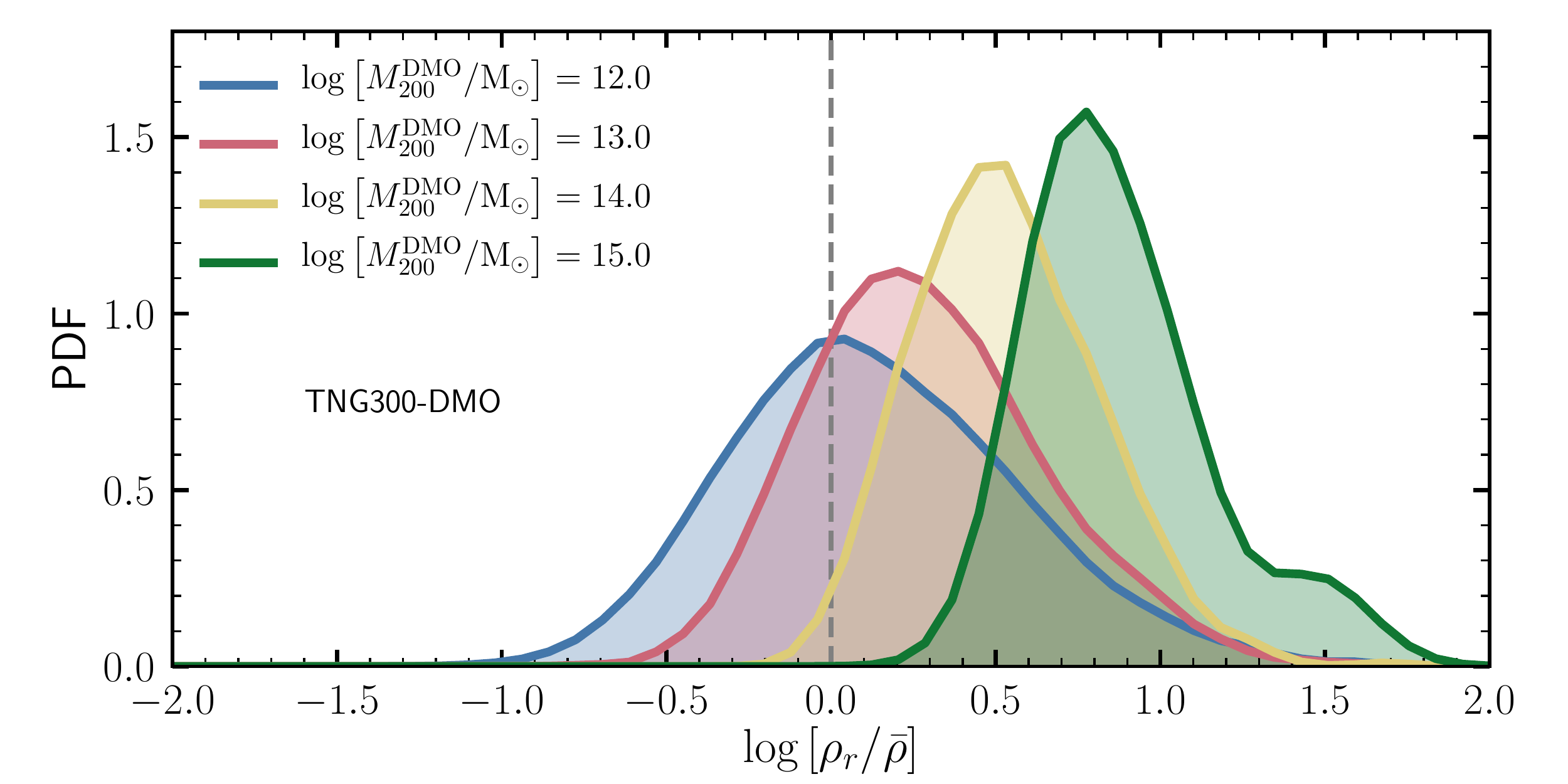}
    \caption{Environmental selection of dark matter haloes from the TNG300 DMO volume. {\it Top left panel}: Histograms of halo environmental overdensity, $\rho_r/\bar{\rho}$, for different aperture radius choices (see text for details). {\it Top right panel}: the locations of haloes existing in the 20 per cent most overdense (red) and 20 per cent most underdense (blue) regions in the TNG300 DMO simulation, defined by the $\rho_r/\bar{\rho}$ metric, assuming a fiducial filter size of $r=5$ Mpc. For clarity, we only display the positions of haloes more massive than $M_{200}^{{\rm DMO}} > 10^{11}\,{\rm M}_\odot$ in a region of size 300 Mpc $\times$ 300 Mpc $\times$ 100 Mpc. {\it Bottom panel}: the distribution of environments occupied by haloes split by mass. Again, we have assumed $r=5$ Mpc.}
    \label{fig:env_def}
\end{figure*}

\begin{enumerate}
    \item {\it Central density}: characterising the central density of a halo provides a quantitative measure for the depth of the potential well that the central galaxy resides in. A popular way to characterise central density is through the concentration parameter, obtained after fitting an analytic profile to the density distribution, $\rho(r)$, in a dark matter halo. Here, we assume the NFW profile \citep{Navarro1996,Navarro1997}:
    \begin{equation}
        \rho (r) = \frac{\rho_s}{\frac{r}{r_s}\left( 1+\frac{r}{r_s} \right)^2}\;,
    \end{equation}
    where $r_s$ and $\rho_s$, respectively, are the scale radius and the  density of the halo at this radius. The concentration is then given by $c_{{\rm NFW}} = r_{200}/r_s$. An unappealing aspect of fitting density profiles to individual haloes is that the procedure requires access to the full particle dataset, not to mention the usual vagaries of profile fitting associated with the choice of radial range the analytic form is fit to, the number of particles in the halo, the minimisation method used etc. We therefore consider an alternative proxy for the central density, the quantity ${\widetilde V_{{\rm max}}} \equiv V_{{\rm max}} / \left(H_0\,r_{{\rm max}}\right)$, where $V_{{\rm max}}$ is the peak of the circular velocity profile of the halo, $r_{{\rm max}}$ is the radius at which this value is attained, and $H_0$ makes this quantity dimensionless. As it is defined, \vmax{} has a one-to-one mapping to the traditional concentration parameter, $c_{{\rm NFW}}$. The benefit of this measure of central density is that the quantities $V_{{\rm max}}$ and $r_{{\rm max}}$ are typically output by all halo finding algorithms, and neither is obtained through fitting.
    
    \item {\it Angular momentum}: the rotation of a halo is best characterised in terms of its (dimensionless) spin parameter, $\lambda$, given by:
    \begin{equation}
     \lambda  = \frac{J \sqrt{\left| E \right|}}{G M_{200}^{5/2}}\;,
    \end{equation}
    \citep{Peebles1969}, where $J$ is the magnitude of the angular momentum of the halo and $E$ is its total energy. While the bulk of the angular momentum of a halo is obtained through tides in the linear density field \citep{Doroshkevich1970,White1984}, the net angular momentum of a halo may be perturbed subsequently through mergers \citep[e.g.][]{Vitvitska2002,Hetznecker2006}. These mergers, in turn, may add to the satellite budget of the halo. We define the spin of a halo using all dark matter particles contained within its $r_{200}$.
    
    \item {\it Epoch of formation}: we classify haloes in the DMO simulation as early or late-forming using their formation redshift, $z_{{\rm form}}$, defined as the epoch by which 50\% of the halo's present day mass is assembled. In practice, this quantity is computed by following the merger tree of each halo along its main progenitor branch; we use merger trees constructed using the {\sc SubLink} algorithm \citep{RodriguezGomez2015}. It is well-established that the redshift of formation and concentration of a halo are correlated, with early-forming haloes exhibiting higher concentrations than their counterparts that collapse later. 
    
    \item {\it Environment}: the large-scale environment in which a halo is embedded may be defined in several ways, with origins rooted in percolation analysis \citep[e.g.][]{Zeldovich1982}, graph theoretical interpretation of the galaxy distribution \citep[e.g.][]{Colberg2007} and by computing tesselations or Hessians of the density field \citep[e.g.][]{Aragon2007,Hahn2007,Sousbie2008,Gonzalez2010,Cautun2013}. A full characterisation of the cosmic web in the TNG simulations is beyond the scope of the present paper; instead, we quantify a halo's large-scale environment using a computationally simpler approach. For each halo in a DMO run, we compute the quantity $\rho_r/\bar{\rho}$, where $\rho_r$ is the mass density in dark matter subhaloes located within a sphere of radius $r$ Mpc from the centre of the halo, while $\bar{\rho}$ is the mean mass density in subhaloes in the entire simulation volume. In computing $\rho_r$, it is important to exclude subhaloes associated with the halo itself (i.e., those within $r_{200}$). Furthermore, we only count subhaloes more massive than $10^9\,{\rm M}_\odot$ in bound dark matter mass. As it is defined, a halo with $\rho_r/\bar{\rho} = 1$ lives at roughly the mean density. The definition adopted here is a mass-weighted version of the one used by \cite{Artale2018}; the advantage of a mass-weighted assignment is that it is capable of distinguishing between pairs of haloes that share the same number of neighbours within a fixed aperture, but where one halo may be located closer to a more massive entity than the other. 

    \end{enumerate}

Fig.~\ref{fig:dmo_props} shows correlations of the central density, spin and formation redshift with halo mass as determined from TNG100-DMO and TNG300-DMO. The top left and top right panels, respectively, show the mass dependence of the central density defined by the NFW concentration of haloes ($c_{{\rm NFW}}$) and our alternative parameterisation, the ratio ${\widetilde V}_{{\rm max}} \equiv V_{{\rm max}}/\left(H_0\,r_{{\rm max}}\right)$. Reassuringly, the latter definition shows a similar dependence on mass as the more traditional mass-concentration relation: low mass haloes have higher concentrations (i.e., higher values of \vmax{}) while the typical scatter in each relation also increases with decreasing halo mass. The mass-concentration relation for the subset of dynamically relaxed haloes is represented by the points with error bars in the top left panel. Here, relaxed haloes are identified according to the criteria in \cite{Neto2007}, which defines relaxed haloes as objects where (i) the mass fraction in substructures contained within $r_{200}$ is less than 10\%, (ii) the offset between the centre of mass of the halo and its centre of potential does not exceed $0.07\,r_{200}$, and (iii) the virial ratio, $2T/\left|U\right| < 1.35$, where $T$ is the kinetic energy of particles within $r_{200}$ and $U$ is their gravitational potential energy. The solid lines in this panel show that one measures systematically lower concentrations in haloes that are out-of-equilibrium (e.g. those that are currently undergoing a merger).

The bottom left panel in Fig.~\ref{fig:dmo_props} shows the median spin-mass relation in TNG. As previous cosmological cold dark matter simulations have shown, there is a very weak (if any) correlation between halo spin and its mass, with a median value of $\lambda\approx0.033$ across a wide range of halo mass \citep[e.g.][]{Davis1985,Barnes1987,Warren1992,Cole1996,Mo1998,Bett2007,Zjupa2017}. Finally, the bottom right panel shows the dependence of the redshift of formation on halo mass, showing the behaviour expected of hierarchical structure formation, with low mass haloes forming earlier than more massive objects. The results from TNG100-DMO and TNG300-DMO are well-converged over the mass scales in which the two simulations overlap; the combination of the simulations enables us to establish these relations over more than four orders of magnitude (although only a limited range is displayed in Fig.~\ref{fig:dmo_props}). 

The top left panel of Fig.~\ref{fig:env_def} shows histograms of halo environmental overdensities as determined by the $\rho_r/\bar{\rho}$ statistic. Similar to \cite{Artale2018}, we have shown the distributions where the radius of the bounding sphere around each halo, $r=3,\,5$ and $8$ Mpc. As expected, the distributions peak around $\rho_r/\bar{\rho}\approx1$, which corresponds to mean density. The top right panel of this figure shows the locations of haloes existing in the 20 per cent most overdense (red) and 20 per cent most underdense (blue) regions in a slab of size 300 Mpc $\times$ 300 Mpc $\times$ 100 Mpc centred on the TNG300-DMO box. In making this projection, we have assumed a filter size of $r=5$ Mpc, which is our fiducial choice hereafter. The topology of filaments and voids is seen clearly using this simple definition for the large-scale environment. Finally, the bottom panel of Fig.~\ref{fig:env_def} shows the distribution of environmental overdensities in which haloes of different mass live. As one would expect na{\"i}vely, more massive haloes exist preferentially in more overdense environments, while lower mass haloes are found in a broader range of overdensities.

\section{Results}
\label{sect:results}

This section presents the main results of this paper. First, we investigate the variation of the HOD as a function of specific DMO properties. In other words: how does the mean number of centrals, $\left<N_{{\rm cen}}\right>$, and satellites, $\left< N_{{\rm sat}}\right>$, hosted by haloes {\it at fixed mass} change when selecting on properties such as concentration, formation time, spin and environment (Section~\ref{sect:hodvars})? We then proceed to investigate the extent to which each of these correlations is independent; this is achieved through the construction of {\it predicted} HOD catalogues (Section~\ref{sect:hodctrl}). Finally, we examine the spatial arrangement of satellite galaxies within haloes, which is another important detail of mock catalogue construction (Section~\ref{sect:radprof}).

\begin{figure*}
    \centering
    \includegraphics[width=0.95\textwidth]{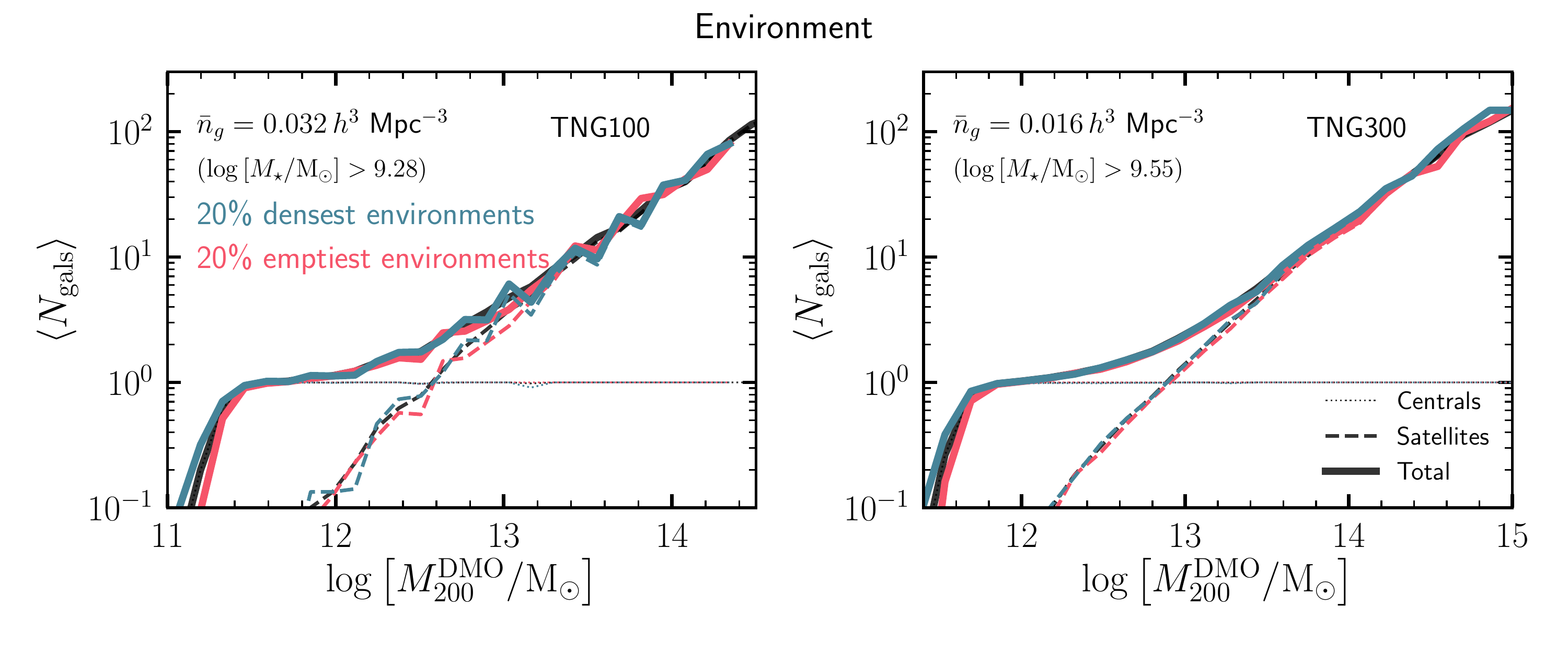}\\
    \includegraphics[width=0.95\textwidth]{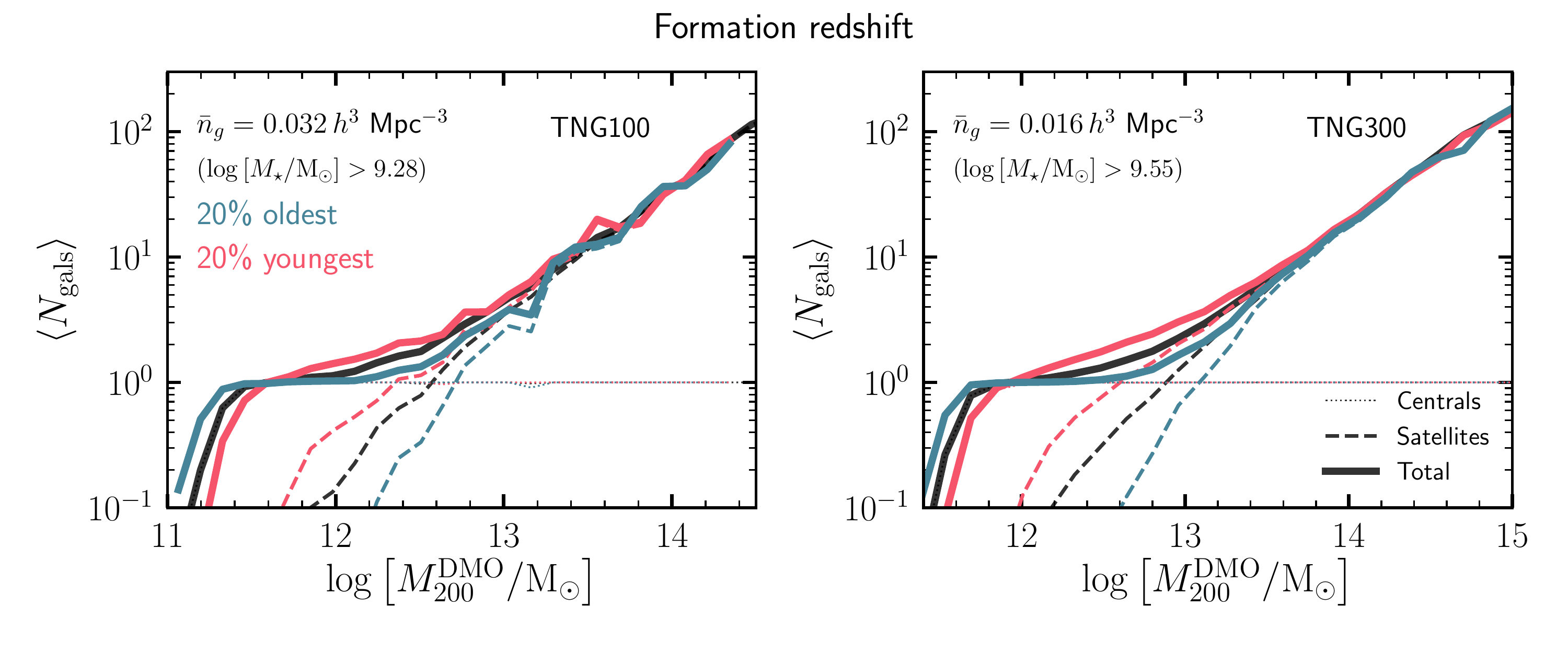}\\
    \includegraphics[width=0.95\textwidth]{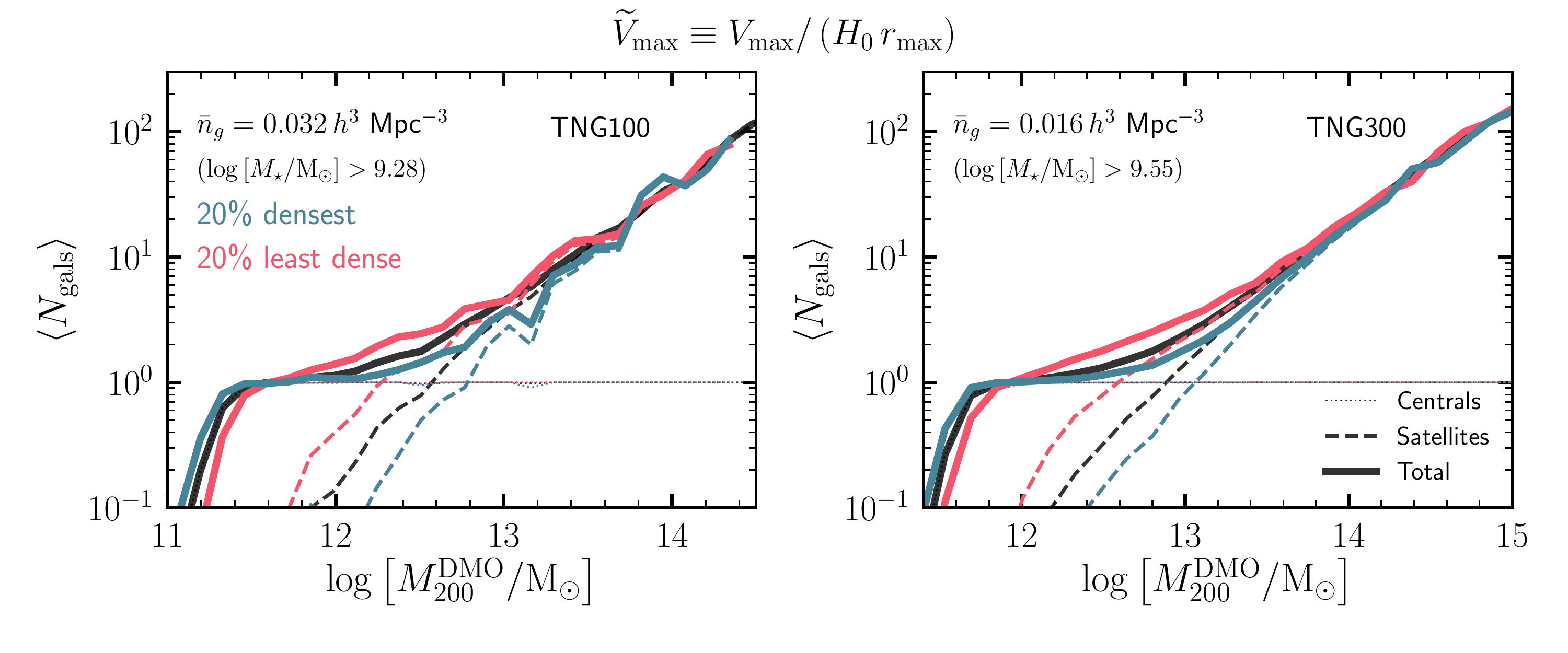}
    \caption{Variation in the mean halo occupation distribution (HOD) in the TNG100 (left column) and TNG300 (right column) simulations. Each HOD is constructed at a fixed number density as indicated in the top left corner of each panel. Each row shows the response of the mean HOD after selecting on various DMO properties: environment, formation redshift and \vmax{}. Centrals are shown in dotted lines, satellites by dashed lines and the total population with solid lines.}
    \label{fig:HOD_vars}
\end{figure*}

\begin{figure*}
    \centering
    \includegraphics[width=0.95\textwidth]{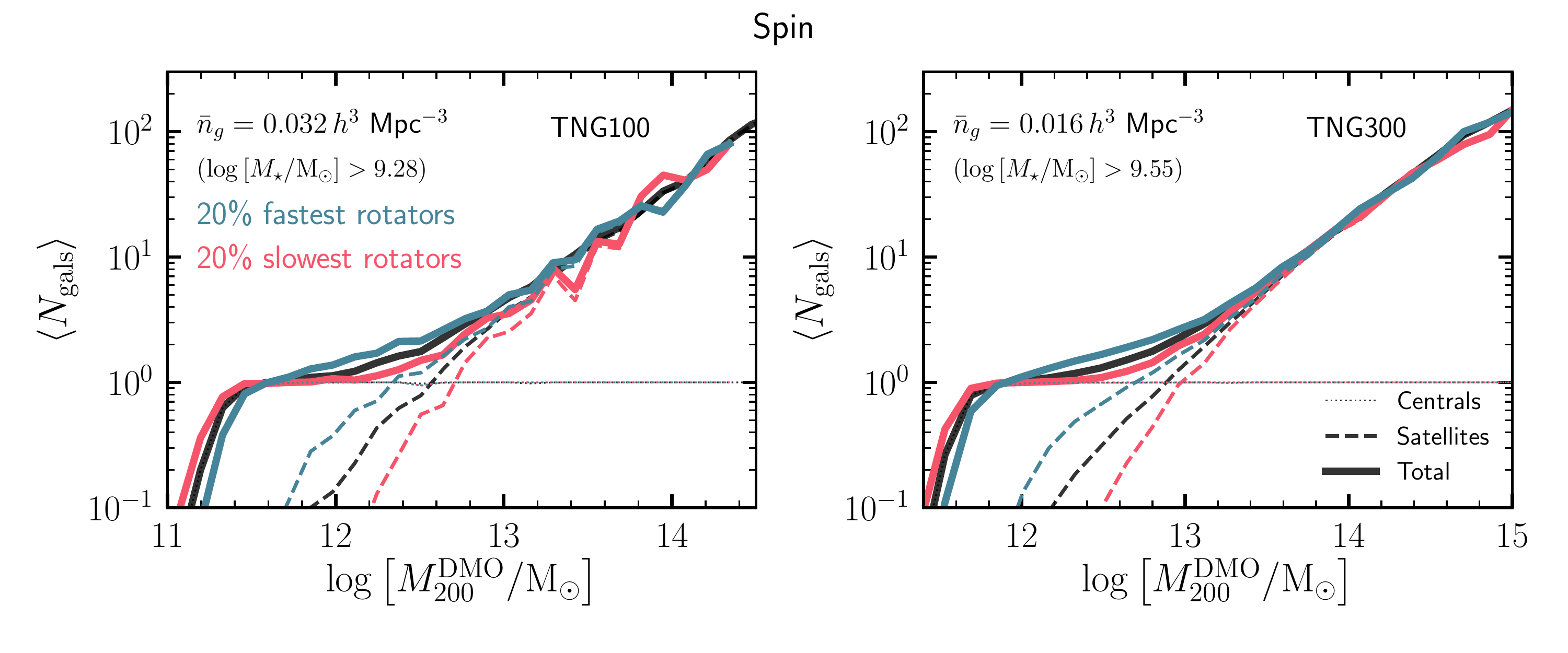}
    \caption{A continuation of Fig.~\ref{fig:HOD_vars} showing variations of the mean HOD with respect to a halo's dimensionless spin parameter.}
    \label{fig:HOD_vars_contd}
\end{figure*}

\subsection{Responses in the Halo Occupation Distribution}
\label{sect:hodvars}

We begin by investigating the response of the HOD to selections on DMO properties at fixed halo mass. In the figures that follow, we showcase these responses by selecting haloes with the 20 per cent highest/lowest values of each DMO property in each bin of halo mass. The results are shown in Figs.~
\ref{fig:HOD_vars} \&~\ref{fig:HOD_vars_contd}, and are summarised below:

\begin{enumerate}
    \item {\underline {\it Environmental bias}}: the top panel of Fig.~\ref{fig:HOD_vars} shows the variation in the mean occupation of galaxies when selecting haloes in the 20 per cent most overdense and 20 per cent most underdense environments (as determined from the corresponding DMO simulation). We find that, at these number densities, the effect of environmental bias is small, particularly in the mean occupation of satellite galaxies. Haloes in the most overdense environments are somewhat more likely to host a central galaxy (with stellar mass $\geq \log M^{\rm min}_\star$) than those in underdense regions. This is signified by the separation of the blue and red curves below the knee of the HOD. This may be attributed to the increased frequency of mergers experienced by haloes in high density environments \citep[e.g.][]{Fakhouri2009}. While the effect on $\left<N_{{\rm sat}}\right>$ is negligible, we note that the extent to which environmental bias introduces scatter in the mean occupation depends on the precise definition of a satellite galaxy i.e., the bounding radius within which a galaxy must be located in order to be counted as a satellite. As described in Section~\ref{sect:matching}, here we have only considered objects located within $r_{200}$; the effect of varying this choice is presented in Appendix~\ref{app:radius}.
    
    \item {\underline {\it Age bias}}: the middle panel of Fig.~\ref{fig:HOD_vars} shows the variation in the mean occupation of galaxies when selecting the 20 per cent earliest-forming and 20 per cent latest-forming haloes in each mass bin. The variation induced by the formation time of haloes is much stronger than environment, predominantly in the mean occupation of satellite galaxies. Younger haloes have, on average, more satellites than older ones; in the former case, satellites will have fallen into their host later and thereby undergo fewer orbital passages (during which they could be destroyed) than satellite galaxies in older haloes. The loss of a satellite may be associated to its bound stars being either transferred into the diffuse starlight in the halo or merged onto the central galaxy, but may also result from a satellite halo being disrupted below our minimum total/stellar mass cut implied by the HOD target density threshold. The relations then cross over around the knee of the HOD (where the contribution of $\left<N_{{\rm sat}}\right>$ becomes sub-dominant to $\left<N_{{\rm cen}}\right>$). At the lowest masses, early-forming haloes are more likely to host a central galaxy through a combination of the added time available to form a central stellar component, and the increased stellar mass deposition from destroyed satellites.
    
    \item {\underline {\it Central density bias}}: the bottom panel of Fig.~\ref{fig:HOD_vars} shows the variation in the mean occupation of galaxies when selecting the 20 per cent most dense and 20 per cent least dense haloes, as determined by the \vmax{}. At all masses, a more concentrated halo has fewer satellites at present day. This suggests that a halo with a higher central concentration of dark matter has a greater propensity to destroy small satellites through tidal forces; as we demonstrate later in this section, these highly concentrated haloes also host more massive central galaxies. As in the case of the variation with formation time, the relations cross over at the knee of the HOD: at low halo mass, a higher concentration halo is on average more likely to host a central galaxy above the minimum stellar mass adopted here ($\log \left[M^{{\rm min}}_\star/{\rm M}_\odot\right]=9.28$ for TNG100 and $\log \left[M^{{\rm min}}_\star/{\rm M}_\odot\right]=9.55$ for TNG300), owing to its deeper potential well. We have checked explicitly that both the trend and the size of this effect is identical when parameterising the central density as the NFW concentration of the halo, $c_{{\rm NFW}}$.
    
    \item {\underline {\it Angular momentum bias}}: finally, Fig.~\ref{fig:HOD_vars_contd} shows the variation in the mean occupation of galaxies when selecting the 20 per cent fastest rotating and 20 per cent slowest rotating haloes, selected according to their values of $\lambda$, the dimensionless spin parameter. The trends with spin may be most readily understood in the context of halo mergers: high angular momentum haloes are predominantly out-of-equilibrium objects that have undergone a relatively recent major merger \citep[e.g.][]{Donghia2007}; these merger events can drag in satellites to add to the existing population within the halo. As we will demonstrate in the following subsection, the dependence of the HOD on halo spin is captured entirely by its dependence on the central density of haloes. 
    
\end{enumerate}
 
\begin{figure*}
    \centering
    \includegraphics[width=0.95\textwidth,height=0.55\textwidth]{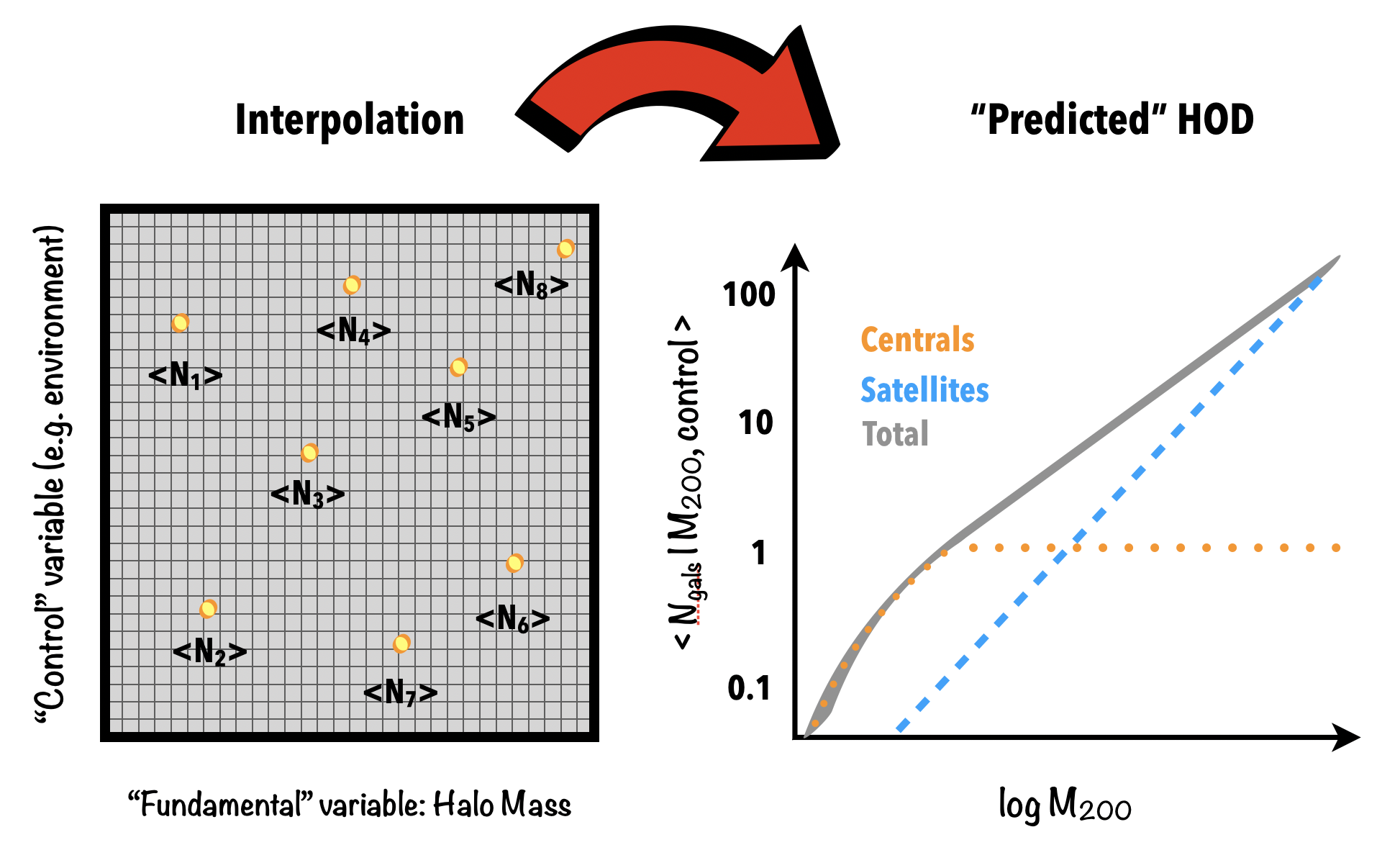}
    \caption{A schematic illustration of how we construct the HOD predicted by pairs of halo properties, by interpolating values of $\left<N_{{\rm cen}}\right>$ and $\left<N_{{\rm sat}}\right>$ on a grid of halo mass and a `control' variable.}
    \label{fig:HOD_pred}
\end{figure*} 
 
The strength of the response in the HOD to each property increases with decreasing halo mass; this is because the typical scatter in environment, formation time, concentration and spin also increases with decreasing halo mass (Figs.~\ref{fig:dmo_props} \&~\ref{fig:env_def}). We note that the dependence on environment, formation time and central density have been established previously, such as in \cite{Zehavi2018} in the context of semi-analytic galaxy formation models, and by \cite{Chua2017} and \cite{Artale2018}, who investigated these correlations using the {\sc Illustris} \citep{Vogelsberger2014a} and {\sc Eagle} \citep{Schaye2015} simulations. The present work examines these relations in the context of the updated TNG model and is able to extend the analysis into the regime of rich clusters using the TNG300 volume. The qualitative response of the HOD to DMO properties is similar in all cases. In the following subsection, we show that each of these responses is not independent and may, in fact, be explained due to underlying correlations between the DMO properties themselves.

\subsection{Independent correlations in the Halo Occupation Distribution}
\label{sect:hodctrl}

In Section~\ref{sect:hodvars}, we have shown that the mean occupation of galaxies in \tng{} responds sensitively to the properties of the dark matter haloes hosting them. While these correlations are interesting physically, they pose a challenge to theoretical models of the HOD if each correlation is independent: to capture the scatter in the galaxy-halo occupancy, it would be necessary to build in the dependence on environment, concentration, formation time etc. in addition to halo mass. The question then arises as to how much weight one should assign to each of these dependencies. 

Of course, each quantity we have examined is not entirely independent of the next. Consider, for example, a halo's large-scale environment. A halo originating from a Lagrangian region located in an overdensity is likely to collapse earlier than one forming in an underdensity. In haloes that collapse earlier, the central core is assembled at an epoch when the mean density of the universe is higher; this results in a larger value of the central density (or concentration) measured at $z=0$ \citep[e.g.][]{Sheth2004,Avila2005,Li2008,GilMarin2011}. These correlations between halo properties, albeit weak, may also propagate into correlations in the HOD. 

We test the independence of the HOD responses observed in Section~\ref{sect:hodvars} through the construction of `synthetic' HODs. First, we construct a grid of dark matter halo mass ($M_{200}^{{\rm DMO}}$, which we label as our `fundamental' variable), a secondary DMO halo property (e.g. environment, which we label as the `control' variable) and the corresponding number of centrals / satellites hosted by this halo in the full physics run (as determined by the matching procedure described in Section~\ref{sect:matching}). Using this space, we construct a spline interpolation function to {\it predict} the average number of centrals / satellites associated with a halo {\it given} its mass and the value of the secondary halo property (e.g. the overdensity it lives in). Mathematically, while the {\it true} HOD measured from \tng{} is given by $\left<N_{{\rm gals}} | M_{200}^{{\rm DMO}} \right>$, the {\it predicted} HOD is defined by $\left<N_{{\rm gals}} | M_{200}^{{\rm DMO}}\,,{\rm control} \right>$. 

\begin{figure*}
    \centering
    \includegraphics[width=0.95\textwidth]{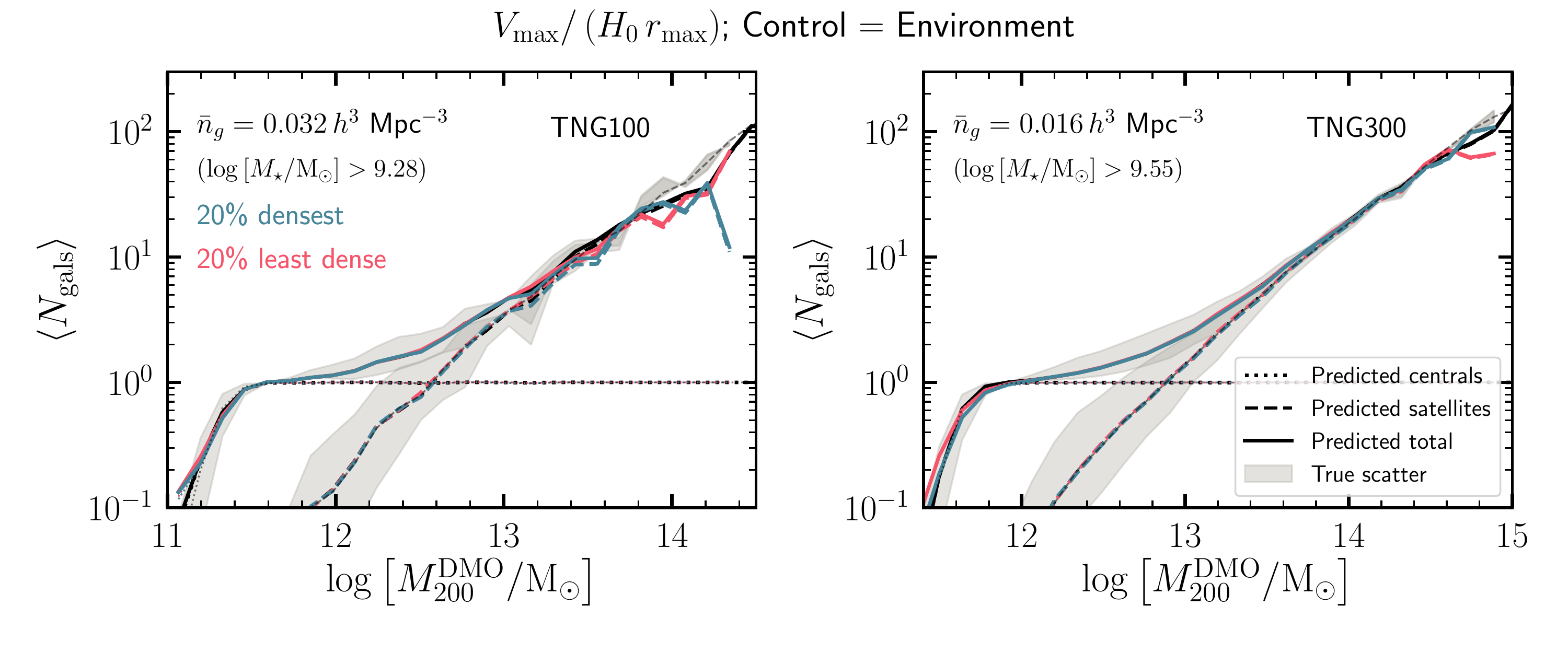}\\
    \includegraphics[width=0.95\textwidth]{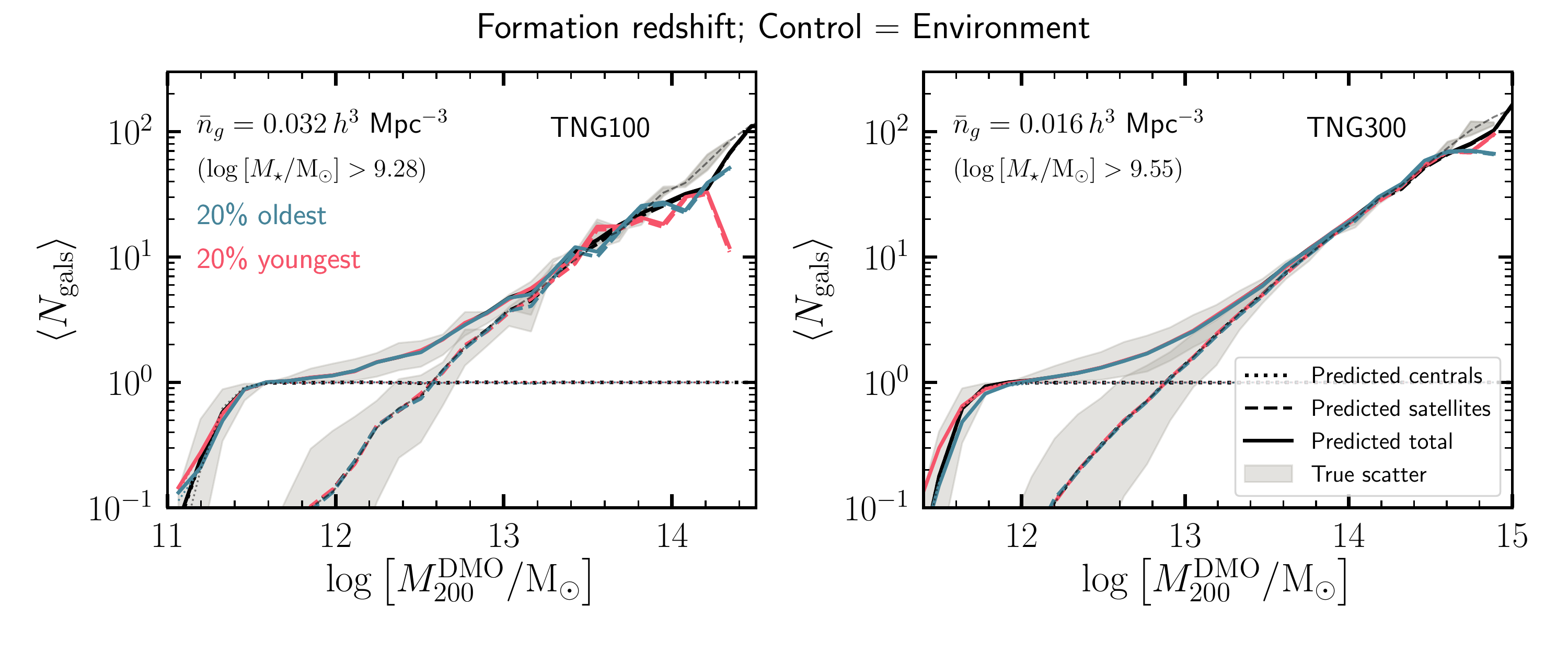}
    \caption{As Fig.~\ref{fig:HOD_vars}, but now showing the mean HOD that is {\it predicted} by a halo's mass, $M_{200}^{{\rm DMO}}$, and its large-scale environment in the DMO simulation. The variation of the HOD with respect to selection on environment is reproduced by construction and is not shown here. Predicting the halo occupancy using halo mass and environment alone is unable to reproduce correlations with other halo properties.}
    \label{fig:HOD_vars_env}
\end{figure*}

Diagrammatically, this procedure is illustrated in Fig.~\ref{fig:HOD_pred}. The new HOD, by construction, reproduces the dependence of the mean occupation on the `control' variable. If after selecting on a {\it third} DMO property we are able to recover the responses observed in Figs.~\ref{fig:HOD_vars} \&~\ref{fig:HOD_vars_contd}, this third parameter does not give us any more information than the bivariate combination of DMO halo mass and DMO `control' variable alone. Conversely, a failure to recover the dependency indicates that there is more information to be gained from including a third parameter in the HOD. The methodology we adopt here is inspired by \cite{Blanton2005}, who investigated the relationship between environment and broadband optical properties like luminosity, surface brightness and colour of galaxies in the Sloan Digital Sky Survey. 

The first control variable we consider is environment. As we have described above, we create grids of $N_{{\rm cen}}\left( M_{200}^{{\rm DMO}}, \rho_r/\bar{\rho}\right)$ and $N_{{\rm sat}}\left( M_{200}^{{\rm DMO}}, \rho_r/\bar{\rho}\right)$. We then consider responses of this new HOD to selection on \vmax{} and formation time; the results are shown in Fig.~\ref{fig:HOD_vars_env}. Clearly, the strong correlations with concentration and formation time we have measured previously are no longer reproduced: this suggests that the combined knowledge of the mass and large-scale environment of a halo is unable to predict the dependence of the HOD on other halo properties. In other words, while pairs of halo properties (e.g. environment and formation time) may themselves be correlated, this does not guarantee that one property is successful at predicting the response of the HOD to the other. This conclusion is akin to the observation made by \cite{Mao2018}, who explored the role of halo properties in determining galaxy clustering. 

\begin{figure*}
    \centering
    \includegraphics[width=0.95\textwidth]{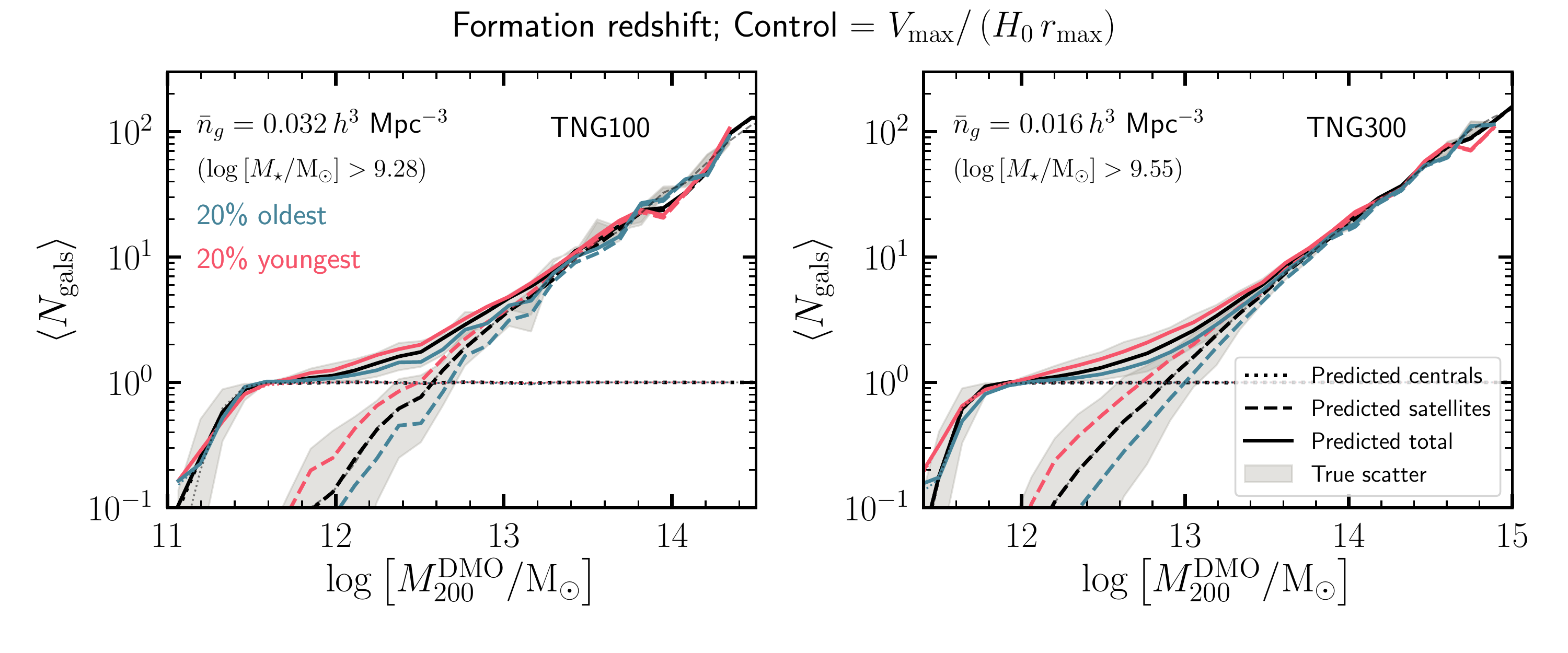}\\
    \includegraphics[width=0.95\textwidth]{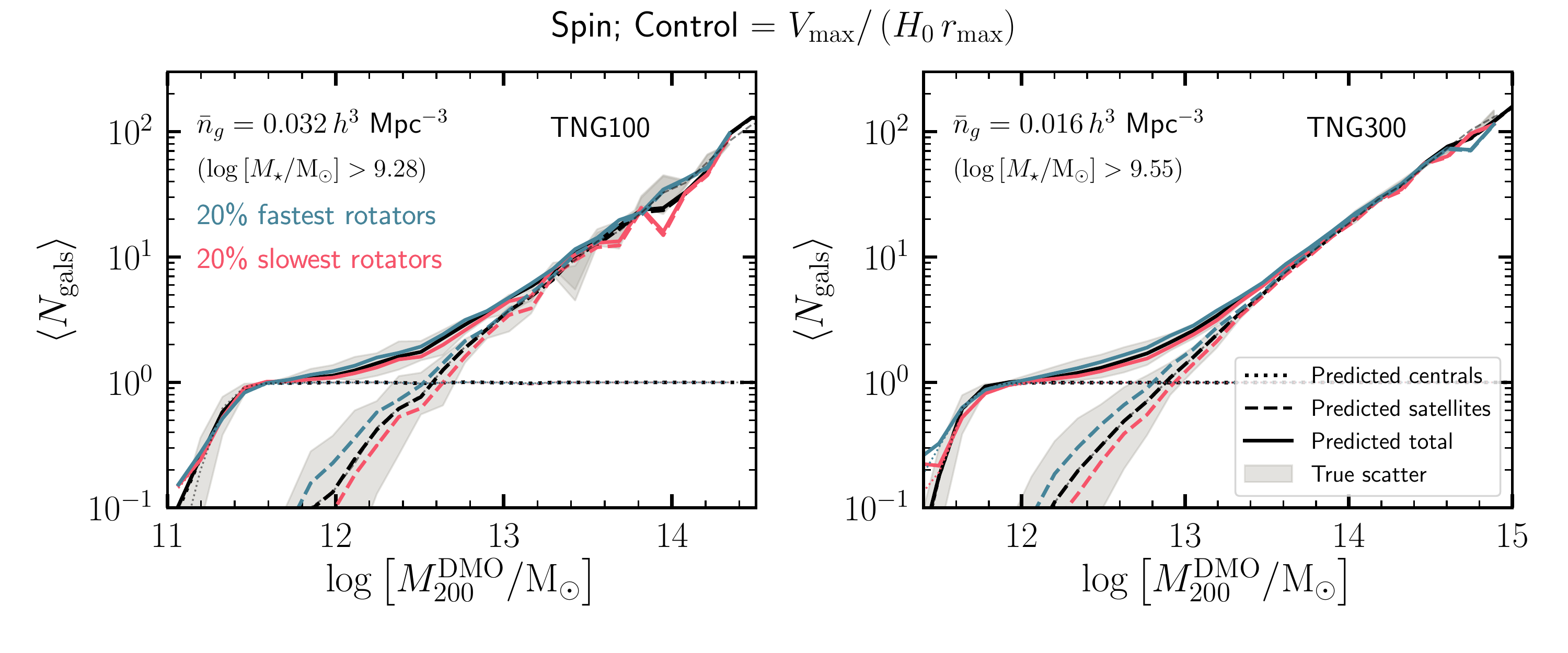} \\
    \includegraphics[width=0.95\textwidth]{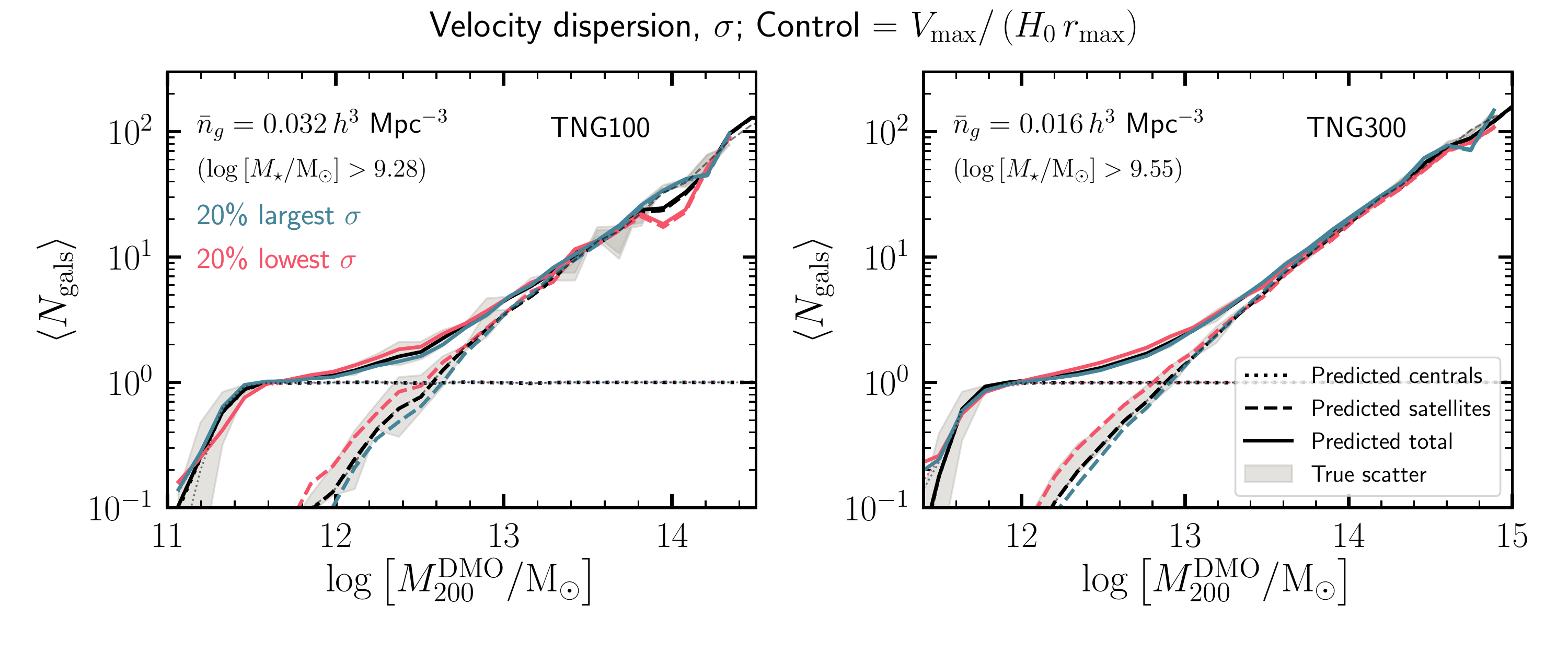}
    \caption{As Fig.~\ref{fig:HOD_vars_env}, but now showing the mean HOD that is {\it predicted} by a halo's mass, $M_{200}^{{\rm DMO}}$, and \vmax{} in the DMO simulation. The variation of the HOD with respect to selection on \vmax{} is reproduced by construction. Velocity dispersion, $\sigma$, shown in the bottom row, is defined as the one dimensional dispersion of dark matter particle velocities contained within $r_{200}$. A predictor based on halo mass and central density is, to a large extent, able to reproduce correlations with other halo properties.}
    \label{fig:HOD_vars_vmax}
\end{figure*}

\begin{figure*}
    \centering
    \includegraphics[width=0.475\textwidth]{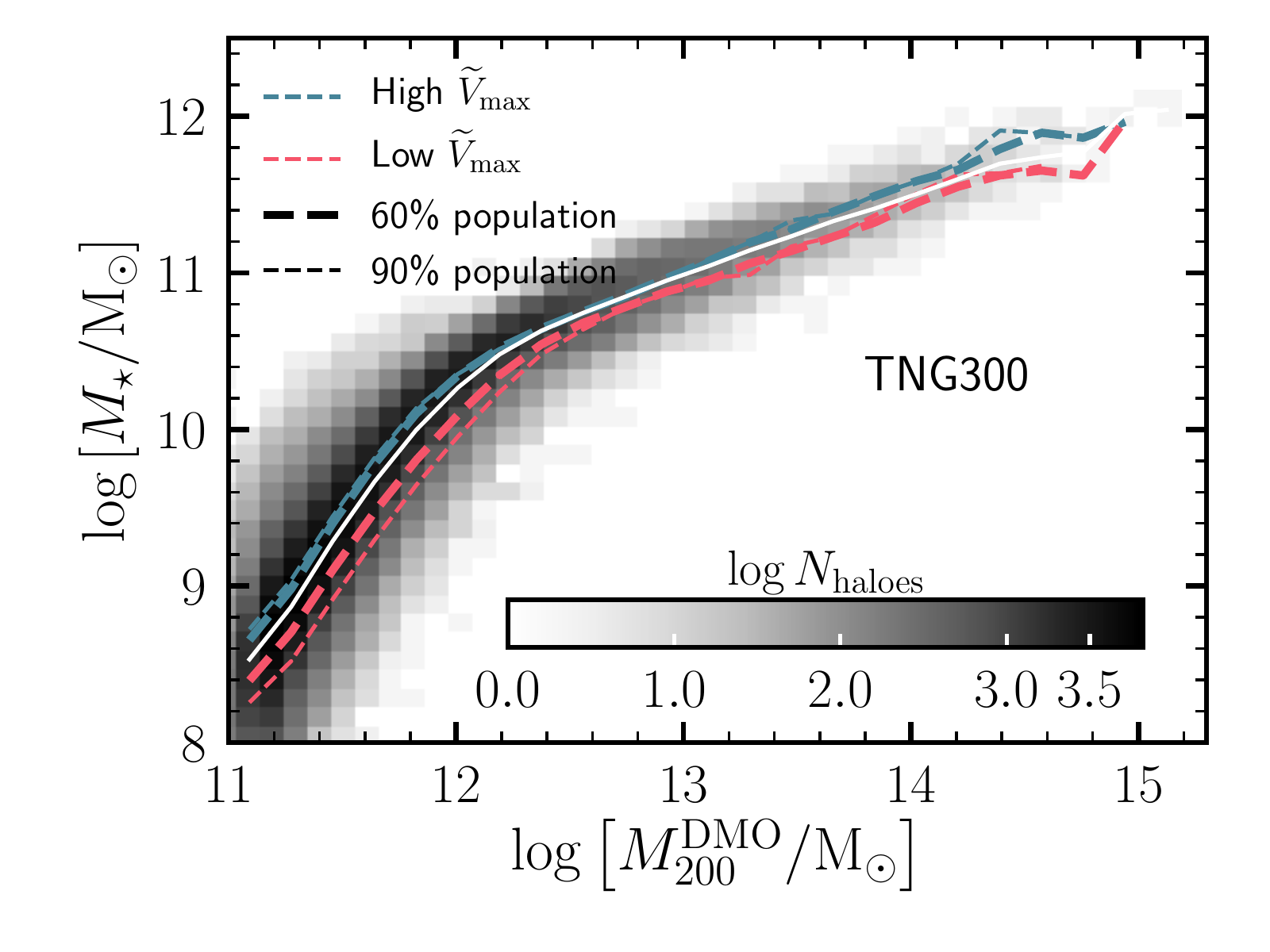}
    \includegraphics[width=0.475\textwidth]{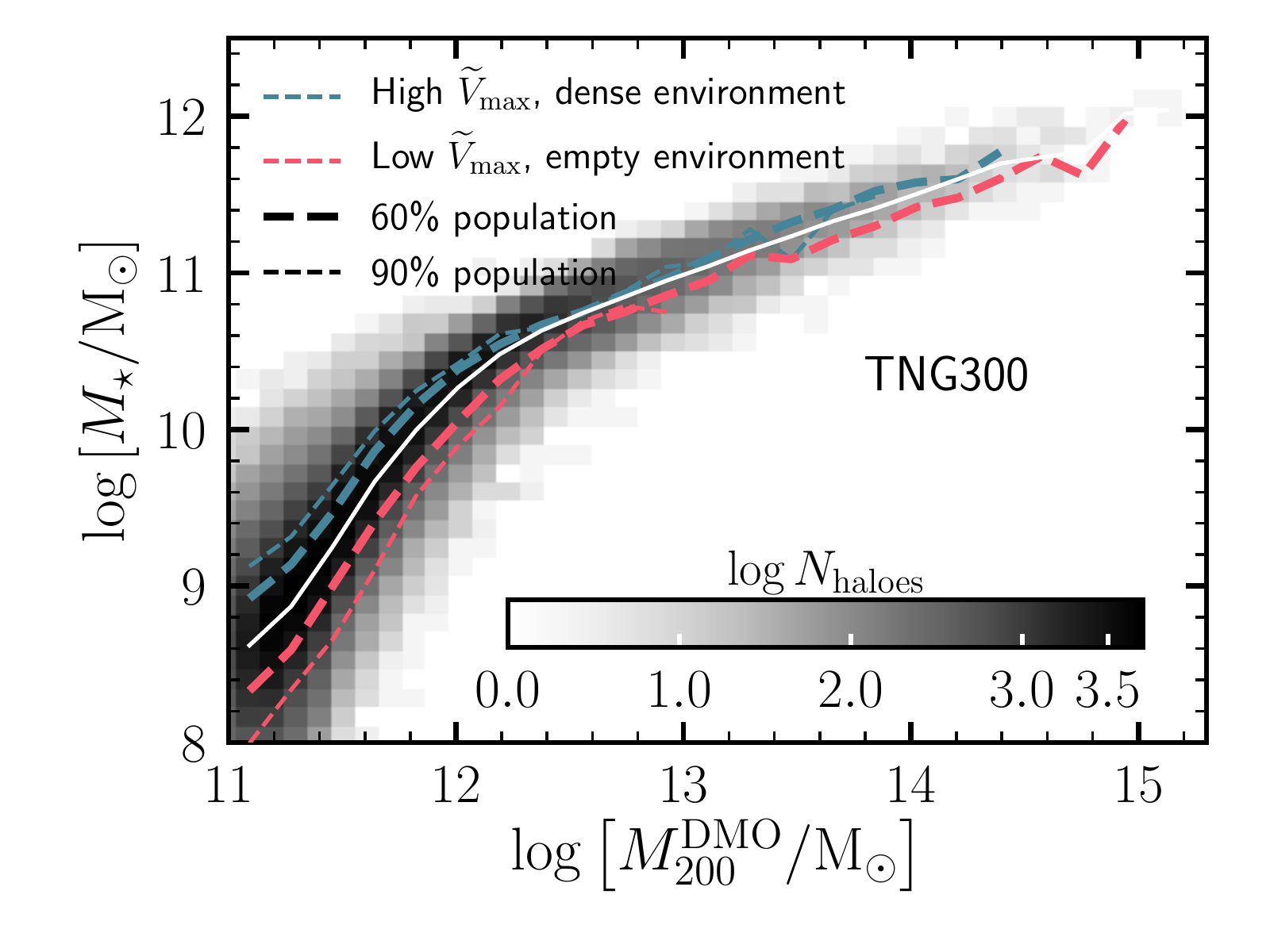}
    \caption{The mean stellar mass, $M_\star$, vs. halo mass, $M_{200}$, relation for central galaxies at $z=0$ in TNG300. The stellar mass of the central galaxy is defined as the total mass in all star particles identified within a 3D aperture of (physical) size 30 kpc placed at the halo centre. {\it Left panel}: The blue and red dashed curves, respectively, show the mean stellar mass-halo mass relation for haloes with higher and lower than average values of \vmax{} at fixed halo mass. The thick dashed curves represent the haloes with the 20 per cent largest (blue) values of \vmax{} and 20 per cent smallest (red) \vmax{} (i.e., encompassing 80 per cent of the population); the thin curves show the same for the 5 per cent and 95 per cent cases (i.e., encompassing 90 per cent of the population). The white curve shows the mean stellar-to-halo mass relation. {\it Right panel}: similar to the panel on the left, but including the environmental overdensity in addition to \vmax{}.}
    \label{fig:stellarHaloMass}
\end{figure*}

The next control variable we select is halo concentration, parameterised by \vmax{}. The HOD predicted by this variable and its associated responses are shown in Fig.~\ref{fig:HOD_vars_vmax}. We now successfully predict the response of the HOD due to formation time, spin and velocity dispersion using halo mass and \vmax{} as the fundamental and control variables, respectively. In particular, the qualitative response of the predicted HOD to each additional property is preserved: for example, early-forming haloes are still predicted to form fewer satellites than late-forming ones, as we inferred from the true HODs measured in TNG. There is some residual scatter, particularly in $\left<N_{{\rm sat}}\right>$, which the ${\widetilde V}_{{\rm max}}$-predicted HODs are unable to capture; this dominates in the regime when only one in every few haloes contain a galaxy above the minimum stellar mass threshold.

 It is worth reflecting about the physical reasons as to why the quantity \vmax{} captures more general dependencies in the HOD. That this parameter also reproduces the dependence on, say, halo spin, is not trivial, but may be understood as follows. At fixed mass, the distribution of high spin haloes are thought to be comprised of systems that have recently undergone a major merger \citep[e.g.][]{Vitvitska2002,Peirani2004,Donghia2007}. These out-of-equilibrium systems are also the ones for which we measure preferentially lower concentrations at fixed mass -- the subset of ``unrelaxed'' haloes represented by the solid lines in Fig.~\ref{fig:dmo_props} (top left panel). High spin and low concentration are therefore intimately related with one another; both predict reduced satellite counts and therefore affect the HOD in the same way. Indeed, on comparing the bottom row of Fig.~\ref{fig:HOD_vars} (concentration) with Fig.~\ref{fig:HOD_vars_contd} (spin), we find that response of $\left<N_{{\rm sat}}\right>$ is similar in both size and shape, but with the colours of the lines reversed.

Ultimately, the efficacy of \vmax{} as a variable for predicting the scatter in the HOD stems from it being a key parameter in quantifying the scatter in the mass of the central galaxy at fixed halo mass. The relationship between the dark matter and stellar content of a halo is a key constraint that galaxy formation models aim to reproduce. Indeed, the {\it mean} stellar-to-halo mass relation at $z=0$, inferred from abundance matching, is one of the quantities that the subgrid physics model in TNG is was designed to reproduce, at least in its general shape \citep{Pillepich2018}. Simulations provide a unique opportunity to explore the scatter around this relation induced by properties intrinsic to haloes, the effects of supernova and AGN feedback etc. \citep[e.g.][]{Trujillo2011,Zentner2014,Matthee2017,He2019}. Such an exploration is performed more easily in the case of central galaxies, which are free of tidal effects from larger host haloes.

Both idealised and fully hydrodynamical simulations have demonstrated that the presence of a massive central galaxy can significantly deplete a halo's satellite population \citep[e.g.][]{Zentner2005,DOnghia2010,Yurin2015,Zhu2016,Sawala2017,GarrisonKimmel2017,Chua2017,Richings2018}. This destruction acts preferentially on low-mass satellites on radial orbits with close pericentric passages to the central disk; the depletion also becomes stronger the more massive the central galaxy \citep{Samuel2019}. Properties of dark matter haloes that capture the scatter in the stellar-to-halo mass relation of central galaxies will naturally be good predictors for the variation in the mean occupancy of satellite galaxies.

In Fig.~\ref{fig:stellarHaloMass} we display the mean and the scatter in the stellar-to-halo mass relation of central galaxies in TNG300\footnote{A comparison between the stellar-to-halo mass relations in TNG and the standard trends inferred from abundance matching has been presented in \citealt{Pillepich2018b}.}. Here, stellar mass is defined as the total mass in star particles contained within a 3D aperture of (physical) size 30 kpc centred on the halo. The dashed lines show shifts in the mean relation (which is shown by the solid white curve) when selecting subsets of the halo population according to their central density only (\vmax{}, left panel) and, in addition, its large-scale environment ($\rho_r/\bar{\rho}$, right panel).

We see clear departures from the mean relation in both panels. In particular, haloes with higher central densities (concentration) host more massive central galaxies; these haloes have deeper potential wells which bring more stellar mass to the centre, and they also form earlier, allowing more time for the central galaxy to build in mass. The right-hand panel of Fig.~\ref{fig:stellarHaloMass} shows that the densest haloes that also exist in most overdense environments host even more massive central galaxies, although the dominant contribution comes from the higher concentration of the halo. The correlations we measure between central stellar mass and concentration are consistent with those measured by \cite{Matthee2017} and \cite{Artale2018}. This dependence naturally translates into the connection between the HOD and the halo concentration. 

We have repeated the exercise of constructing HODs predicted with other DMO properties, using spin, formation time and velocity dispersion as control variables. Neither spin nor velocity dispersion are as effective as \vmax{} in recovering correlations in the HOD with a third variable. The formation redshift, $z_{{\rm form}}$, is the next best performing parameter. For more generalised applications, however, \vmax{} is a more convenient second parameter to incorporate into an HOD as it is well-defined: to compute this quantity, there is no need to traverse merger trees in which the number of outputs used to construct the tree affects the accuracy with which $z_{{\rm form}}$ may be measured. Parameterising central density as \vmax{} as opposed to the concentration of an equivalent NFW halo also circumvents the need to access particle data. 

\subsection{The radial distribution of galaxies in haloes}
\label{sect:radprof}

\begin{figure*}
    \centering
    \includegraphics[scale=0.7]{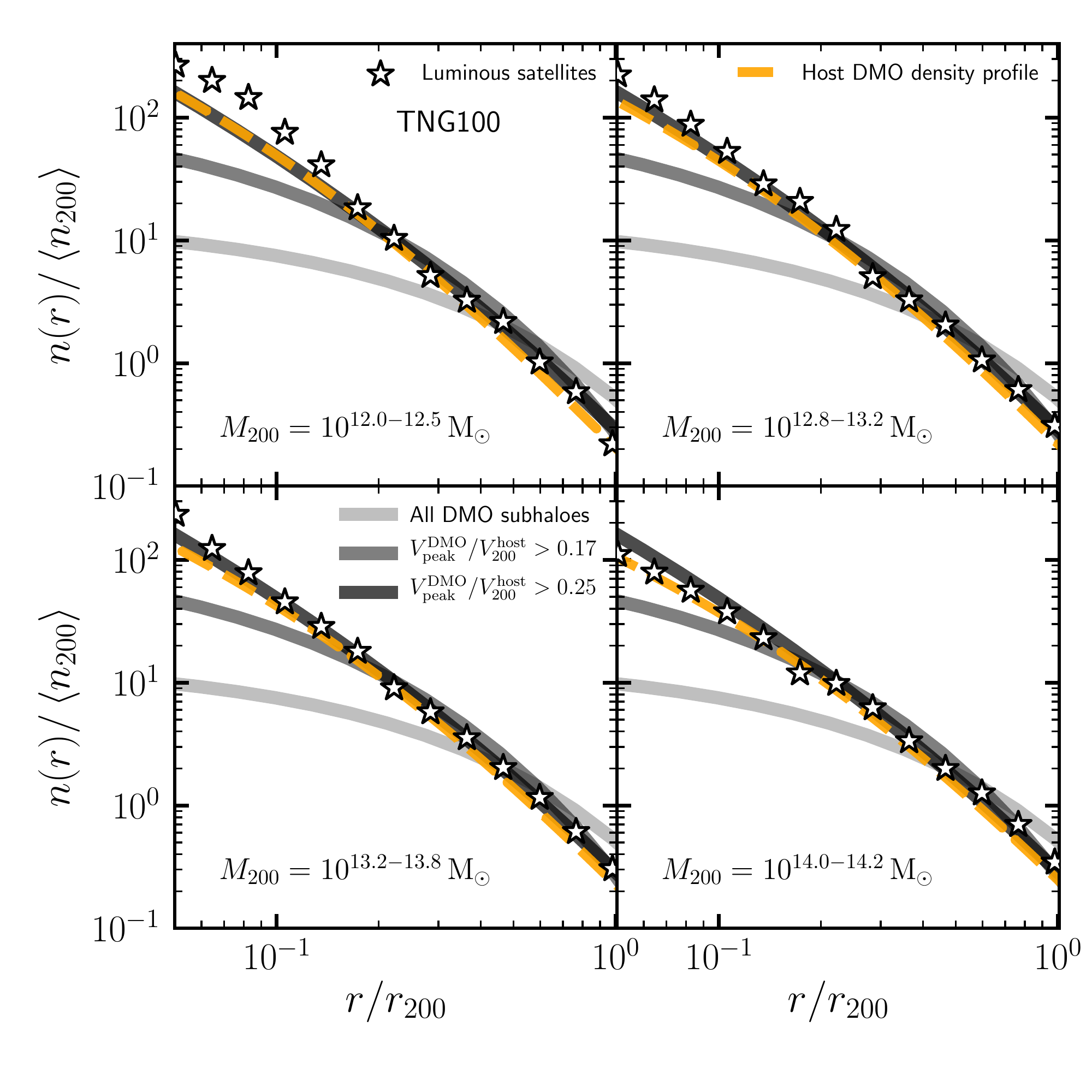}
    \caption{The mean radial number density of luminous satellites normalised to the mean number of luminous galaxies in bins of host halo mass (individual panels). The light grey curve shows the normalised radial profile of dark matter subhaloes contained within hosts of this mass in the DMO simulation; this is a mass-selected sample with the requirement that each subhalo contains at least 100 dark matter particles at present day. The dark grey and black curves show radial profiles of subhalo populations selected according to $V_{{\rm peak}}^{{\rm DMO}}/V_{200}^{{\rm host}}$, the ratio of the subhalo's peak circular velocity to the virial velocity of its host. The $V_{{\rm peak}}$-selected profiles do not include any $z=0$ particle number resolution limit. Finally, the dashed orange curve shows an NFW profile fit to the density profile of DMO haloes in this mass range. For clarity, we have display the results for the TNG100 simulation only; radial profiles in the TNG300 simulation are qualitatively similar.}
    \label{fig:sat_rad_distr}
\end{figure*}

We conclude our analysis by examining how satellite galaxies in \tng{} are distributed spatially within their host haloes. In the HOD formalism, an appropriate assignment of galaxy positions is vital to recover accurate small-scale ($\lesssim 1$ Mpc) clustering (the so-called `one-halo' term). Typically, satellite positions are assigned in one of three ways: (i) assuming that satellites trace the dark matter particles and are distributed according to the best-fitting NFW profile of the host halo; (ii) by assigning satellite positions by placing each one on a randomly-selected dark matter particle; (iii) assuming they follow the radial distribution of subhaloes in a DMO simulation. In a hydrodynamical simulation like \tng{}, the orbits of satellite galaxies are evolved self-consistently, providing knowledge of this assignment for free.

In Fig.~\ref{fig:sat_rad_distr} we present the radial profiles of satellite galaxies measured in TNG100, which offers better mass and spatial resolution than TNG300. In particular, we show the radial number density of luminous galaxies (containing at least 100 star particles), $n(r)$, normalised to the mean number density of satellites identified within $r_{200}$, denoted as $\left<n_{200}\right>$. These profiles are represented by the starred symbols, with each panel showing the result for a different range of host halo mass. The normalised profiles take on a universal shape across all halo mass; their shape and normalisation is independent of the minimum stellar mass cut adopted in constructing the satellite profiles (see Appendix~\ref{app:partprof}. The light grey curve in each panel shows the quantity $n(r)/\left<n_{200}\right>$ measured for subhaloes (containing more than 100 dark matter particles) in the DMO counterparts of these host haloes. The distribution of subhaloes from TNG100-DMO is markedly different to the distribution of galaxies, which is much more centrally concentrated. It is well-known that the distribution of subhaloes in a DMO simulation may be fit with an Einasto profile \citep{Einasto1965}:
\begin{equation}
    \log \left[ \frac{n(r)}{n_{-2}} \right] = -\frac{2}{\alpha} \left[ \left( \frac{r}{r_{-2}} \right)^\alpha - 1 \right]\;,
\end{equation}
with a shape parameter, $\alpha=0.678$, and the parameters $n_{-2}$ and $r_{-2}$ obtained through fitting \citep[e.g.][]{Springel2008}. Furthermore, this shape is independent of subhalo mass \citep[e.g.][]{Gill2004,DeLucia2004,Gao2004}. As shown in Fig.~\ref{fig:sat_rad_distr}, an Einasto profile fit to DMO subhaloes is clearly inaccurate when applied to galaxies.

The dashed orange curve shows an NFW profile where the concentration is set to the median concentration of DMO haloes in this mass bin, and, perhaps surprisingly, provides an excellent match the radial distribution of satellite galaxies in TNG100. To understand why galaxies tend to prefer the more centrally concentrated profile of the total matter, as opposed to the shallower profile of the DMO subhaloes, it is important to identify the subset of subhalo population that is most likely to host a galaxy. For this it is useful to think about the galaxy-halo connection in terms of subhalo abundance matching. In its most traditional form, abundance matching associates the brightest galaxies with the most massive subhalo, defined by values measured at $z=0$. Unlike independent haloes, subhaloes (the hosts of satellite galaxies) are subject to tidal stripping and therefore lose mass continually after infall by an amount that depends on their orbit, infall mass and infall time: subhaloes that are more massive prior to infall (and therefore more likely to host a galaxy) undergo greater dynamical friction and sink to the middle of halo, where the mass is stripped more rapidly. Information about the mass being high before infall may therefore be lost through dynamical evolution. Furthermore, substructure finding algorithms are likely to underestimate the mass of subhaloes near the centre of the host halo, or may even fail to detect them altogether.  

A more powerful proxy of the potential well associated with a subhalo is provided the quantity $V_{{\rm peak}}^{{\rm DMO}}$, defined as the maximum value of $V_{{\rm max}}$ attained by a DMO subhalo at {\it any point} in its history (typically just before infall). This quantity, free from the effects of stripping provides a more direct connection between a halo and its galaxy and has indeed been demonstrated to be a more robust metric for subhalo abundance matching \citep[e.g.][]{Conroy2006,Trujillo2011,Nuza2013,Chaves2016}.

Selecting subhaloes by $V_{{\rm peak}}^{{\rm DMO}}$ rather than present-day mass results in considerably different radial distributions, as shown by the dark grey and black curves in Fig.~\ref{fig:sat_rad_distr}. Subhaloes in TNG100-DMO with the highest values of $V_{{\rm peak}}^{{\rm DMO}}$ are more centrally concentrated; objects with $V_{{\rm peak}}^{{\rm DMO}}/V_{200}^{{\rm host}} > 0.25$ are sorted nearly identically to the total matter distribution. Note that we do not impose any $z=0$ particle number cut when constructing the $V_{{\rm peak}}$-selected radial profiles. These observations are consistent with the findings of previous works by \cite{Nagai2005}, \cite{Faltenbacher2006}, \cite{Kuhlen2007} and \cite{Ludlow2009}. As it has been shown by \cite{Springel2008}, the spatial bias of subhalo clustering is camouflaged when these objects are selected by their $z=0$ mass. 

In summary, we find that the radial profiles of galaxies in TNG are described very accurately by the best-fitting NFW profile of the dark matter particles in a DMO simulation. This represents the centrally-concentrated distribution of subhaloes with the highest values of $V_{{\rm peak}}^{{\rm DMO}}$, which are the entities that are most likely to host a galaxy at present day. On the other hand, the radial profile of DMO subhaloes selected by present-day mass is much shallower than that of galaxies, particularly within $r\lesssim0.4r_{200}$. The correct assignment of galaxy positions within their host haloes influences the amplitude of the one-halo term in the two-point correlation function \citep[see e.g.][]{Reddick2013}.

\section{Conclusions}
\label{sect:conclusions}

In this paper we have presented a detailed study of the galaxy-halo connection as revealed by a modern cosmological, hydrodynamical simulation. In particular, we exploit the combination of high mass resolution and large volume of the 100 and 300 Mpc boxes produced as part of the \tng{} project \citep{Pillepich2018b,Nelson2018a,Naiman2018,Marinacci2018,Springel2018} to explore the Halo Occupation Distribution (HOD). 

Our objective has been to identify properties of haloes in a dark matter-only (DMO) simulation that are most predictive of its likelihood to host central / satellite galaxies in a counterpart hydrodynamical simulation. We achieve this by matching haloes more massive than $\sim10^{11}\,{\rm M}_\odot$ between DMO and full physics versions of each TNG volume, after which we construct galaxy catalogues at fixed number density whilst being careful to include only those galaxies that are well resolved (Figs.~\ref{fig:smf} \&~\ref{fig:HOD_fits}). In TNG100, we construct the HOD for a number density, $\bar{n}_g = 0.032\,h^3$Mpc$^{-3}$ and $\bar{n}_g = 0.016\,h^3$Mpc$^{-3}$ in TNG300; we provide fitting formulae for the HODs resulting from these number densities in Eqs.~\ref{eq:ncen},~\ref{eq:nsat} and Table~\ref{tab:HOD_fit_params}. The minimum galaxy stellar mass implied by these number density thresholds  is $\log\, [M^{\rm min}_\star/{\rm M}_\odot] = 9.28$ in TNG100 and $\log\, [M^{\rm min}_\star/{\rm M}_\odot] = 9.55$ in TNG300. We focus specifically on the following properties and their influence on the mean HOD: concentration (parameterised as the quantity ${\widetilde V}_{{\rm max}} \equiv V_{{\rm max}}/\left(H_0\, r_{{\rm max}}\right)$), formation time, spin (Fig.~\ref{fig:dmo_props}) and environment (Fig.~\ref{fig:env_def}). Our findings are summarised below:
\begin{enumerate}
    \item The response of the HOD is very weakly correlated with environment, although haloes in high density environments are somewhat more likely to host more satellites than those living in underdensities. Low-mass haloes are more likely to host a central galaxy (with stellar mass $\geq \log M^{\rm min}_\star)$ if located in a high density region (Fig.~\ref{fig:HOD_vars}, top panel). The extent to which environmental bias plays a part in introducing scatter to the HOD depends on the outermost boundary used to define the satellite content of a halo (Appendix~\ref{app:radius}).
    \item Younger haloes have, on average, more satellites than older ones; these satellites have had more time to orbit the halo and potentially merge onto the central galaxy (Fig.~\ref{fig:HOD_vars}, middle panel).
    \item At all masses, a more concentrated halo has fewer satellites at present day. At low halo mass, a higher concentration halo is on average more likely to host a central galaxy, owing to its deeper potential well (Fig.~\ref{fig:HOD_vars}, bottom panel).
    \item Haloes with larger angular momentum contain more satellites than the average population, and vice-versa. Previous works have indicated that high angular momentum haloes have typically undergone a recent merger event; these out-of-equilibrium haloes exhibit lower concentrations and later formation times (Fig.~\ref{fig:HOD_vars_contd}). \\
To test the independence of the HOD response to each variable, we use pairs of halo properties to construct `controlled' galaxy catalogues: where the mean occupation of centrals and satellites are {\it predicted} using a combination of halo mass and a second, control variable (Fig.~\ref{fig:HOD_pred}). These tests show that:
    \item Despite intrinsic correlation between environment and halo formation time / concentration / spin, an HOD predicted using halo mass and large-scale environment alone is unable to capture the correlations with other halo properties (Fig.~\ref{fig:HOD_vars_env})
    \item Among the remaining properties, halo concentration as defined by \vmax{} performs the best in predicting the response of the HOD to tertiary properties like environment, spin and velocity dispersion (Fig.~\ref{fig:HOD_vars_vmax}). Formation time performs almost as well, with the disadvantage that it is not as easily determined using $z=0$ halo properties.
    \item The success of concentration as a secondary variable in the HOD is driven by the fact that haloes with steeper inner potentials are more efficient at destroying orbiting satellites. Furthermore, concentration acts as a key secondary parameter in describing the scatter of central galaxy mass at fixed halo mass: haloes with higher concentration host, on average, a more massive central galaxy than those with lower concentration (Fig.~\ref{fig:stellarHaloMass}). A more massive central galaxy may have a higher propensity to destroy orbiting satellites. Furthermore, a satellite orbiting an early-forming, high concentration host halo has more time to merge onto the central galaxy by $z=0$.
    \item Finally, we measure the radial distribution of luminous satellite galaxies in TNG haloes and find that their profile is described very well by an NFW profile fit to the total matter distribution in the DMO counterparts to these haloes (Fig.~\ref{fig:sat_rad_distr}). This reflects the more centrally-concentrated distribution of subhaloes with the highest values of peak circular velocity, $V_{{\rm peak}}^{{\rm DMO}}$, which these satellites occupy preferentially. On the other hand, a sample of DMO subhaloes selected by present-day mass exhibits a comparatively shallower profile towards the centre of the host halo. 
\end{enumerate}
The HOD formalism provides a convenient framework to understand how the simplest property of a dark matter halo -- namely, its mass -- may be used to predict the probability for it to contain central and/or satellite galaxies. Cosmological, hydrodynamical simulations that model simultaneously the evolution of dark and baryonic matter provide the ideal test bed for exploring and quantifying the scatter in the HOD as a function of secondary halo properties. 

In TNG, we find that halo mass and concentration, defined as the quantity ${\widetilde V}_{{\rm max}} \equiv V_{{\rm max}}/\left(H_0\,r_{{\rm max}} \right)$, are two particularly prominent variables in determining the occupancy of galaxies in haloes. This combination of parameters is especially enticing as each one is standard output of all halo finding algorithms, thereby circumventing the need to fit density profiles to particle data or traverse merger trees. The physical quantities that these parameters represent are general enough that they may be readily used to construct HOD catalogues for much larger volume simulations as demanded by the next generation of galaxy surveys for which equivalent hydrodynamical simulations will not be computationally feasible. Finally, since halo mass, $V_{{\rm max}}$ and $r_{{\rm max}}$ evolve self-consistently according to the underlying theory of gravity, the parameterisation we have explored may also be applied to cosmological models beyond traditional $\Lambda$CDM.


\section*{Acknowledgements}
SB is supported by Harvard University through the ITC Fellowship. FM acknowledges support through the Program ``Rita Levi Montalcini'' of the Italian MIUR. The flagship simulations of the IllustrisTNG project used in this work (TNG100 and TNG300) have been run on the HazelHen Cray XC40-system at the High Performance Computing Center Stuttgart as part of project GCS-ILLU of the Gauss centres for Supercomputing (GCS).

\appendix
\section{Environmental assembly bias and the definition of a satellite}
\label{app:radius}

In Fig.~\ref{fig:HOD_vars}, we showed that while the HOD does indeed respond to environmental overdensity, the correlation is weak. In this Appendix we display the relation between the HOD and halo environment for different choices of the boundary used to define the extent of a halo. 

\begin{figure}
    \centering
    \includegraphics[width=\columnwidth]{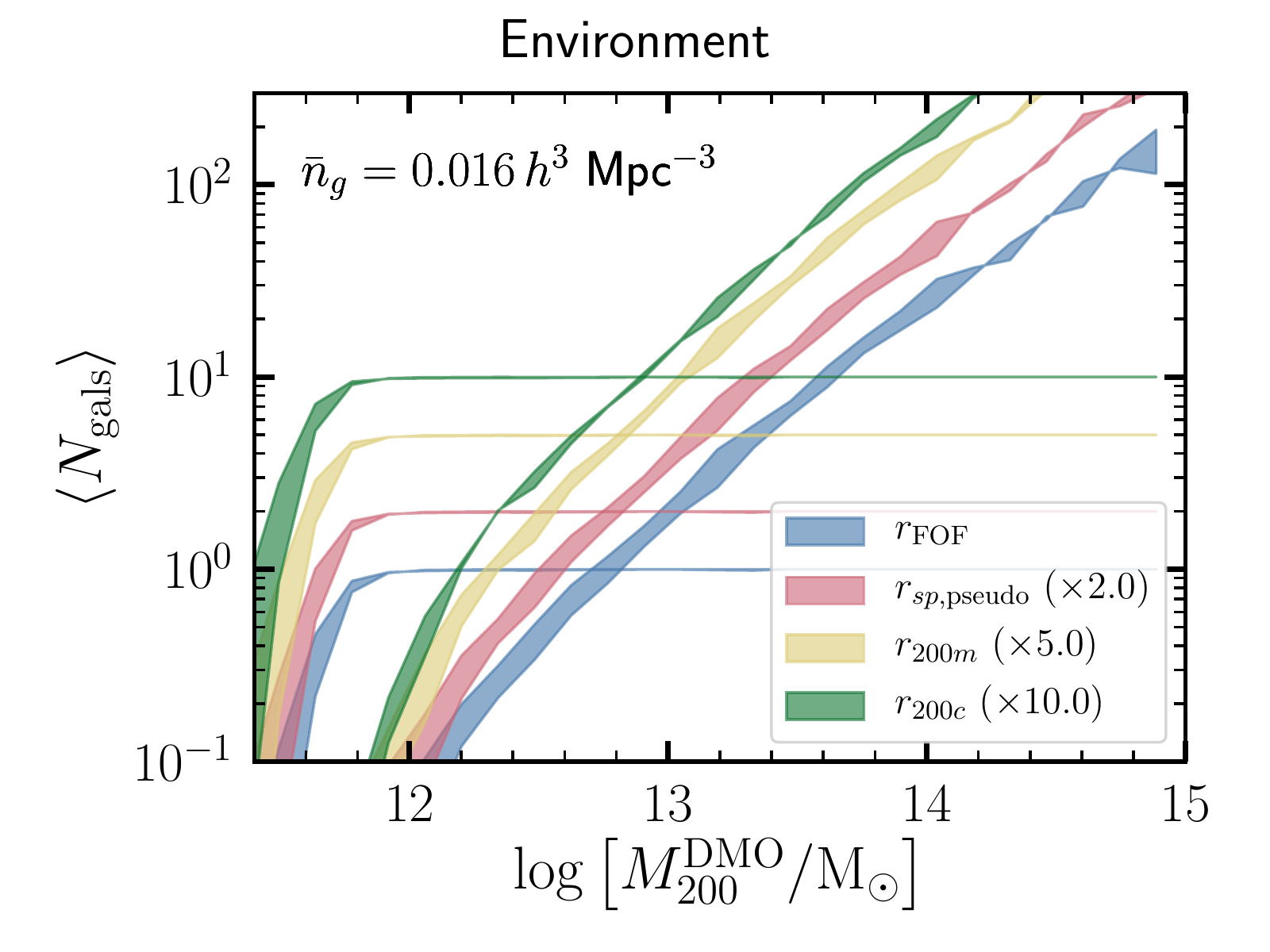}
    \caption{The effect of environment on the HOD for different choices of the maximal radius used to define a satellite galaxy. $r_{200c}$ ($r_{200m}$),  corresponds to the radius within which the mean density of the halo is 200 times the critical (mean) density of the universe; satellites of a halo are defined as the set of galaxies located within this boundary. $r_{{\rm FOF}}$ refers to the scenario in which a satellite is any galaxy located within the FOF group as a whole. The definition of halo environment for each choice of boundary is adjusted accordingly (see Section~\ref{sect:haloprops} for details). Finally, $r_{sp,{\rm pseudo}}$ is a proxy for the splashback radius, which we have assumed to correspond to $1.5\,r_{200m}$. Here, we show the results for TNG300 only.}
    \label{fig:HOD_radvars}
\end{figure}

Fig.~\ref{fig:HOD_radvars} shows the response of the HOD to halo environment (i.e., similar to the top row of Fig.~\ref{fig:HOD_vars}) where the boundary of a halo is chosen to be $r_{200c}$ (our fiducial choice, where the enclosed overdensity is equal to 200 times the critical density), $r_{200m}$ (enclosed density is 200 times the mean density) and $r_{{\rm FOF}}$ (which simply selects the irregular edge defined by dark matter particles linked by the friends-of-friends algorithm). In general, $r_{200c} < r_{200m} < r_{{\rm FOF}}$. A satellite galaxy in each catalogue is defined as a galaxy that is located within the corresponding definition of the host halo's boundary, ignoring the {\sc subfind}-based central vs. satellite classifications. The quantity $\rho_r/\bar{\rho}$, which characterises the environmental overdensity, is adjusted to be consistent with each definition (see Section~\ref{sect:haloprops} for details). The shaded regions in the figure encompass the variation in the mean occupancy of centrals and satellites for haloes located in the 20 per cent most overdense and 20 per cent most underdense environments -- this is simply the area enclosed between the red and blue curves in Fig.~\ref{fig:HOD_vars}. The curves corresponding to $r_{200m}$ and $r_{200c}$, respectively, have been multiplied by factors of 5 and 10 to facilitate comparison.

It is clear from Fig.~\ref{fig:HOD_radvars} that the extent to which environmental bias manifests in the HOD depends on the choice of halo boundary. This is true particularly in the case of the mean satellite occupation, which varies more strongly with environment for larger boundary definitions. This highlights the importance of a physically-motivated choice for the boundary of a halo, such as the so-called `splashback' radius \citep[e.g.][]{Fillmore1984,Bertschinger1985,Adhikari2014,Diemer2014,More2015}. In Fig.~\ref{fig:HOD_radvars}, we use $r_{sp,{\rm pseudo}}=1.5\,r_{200m}$ as a rough proxy for this radius \citep{Diemer2014}. The variation of the mean central occupation, on the other hand, is similar for all boundary definitions.

\section{Numerical considerations in satellite radial profiles}
\label{app:partprof}

In TNG, we find that the normalised radial distribution of galaxies, $n(r)/\left<n_{200}\right>$, follows a universal profile that is described very well by the best-fitting NFW profile of their host haloes in the DMO simulation (Fig.~\ref{fig:sat_rad_distr}). This Appendix establishes that this observation is independent of the mass range of satellite galaxies. 

\begin{figure}
    \centering
    \includegraphics[width=\columnwidth]{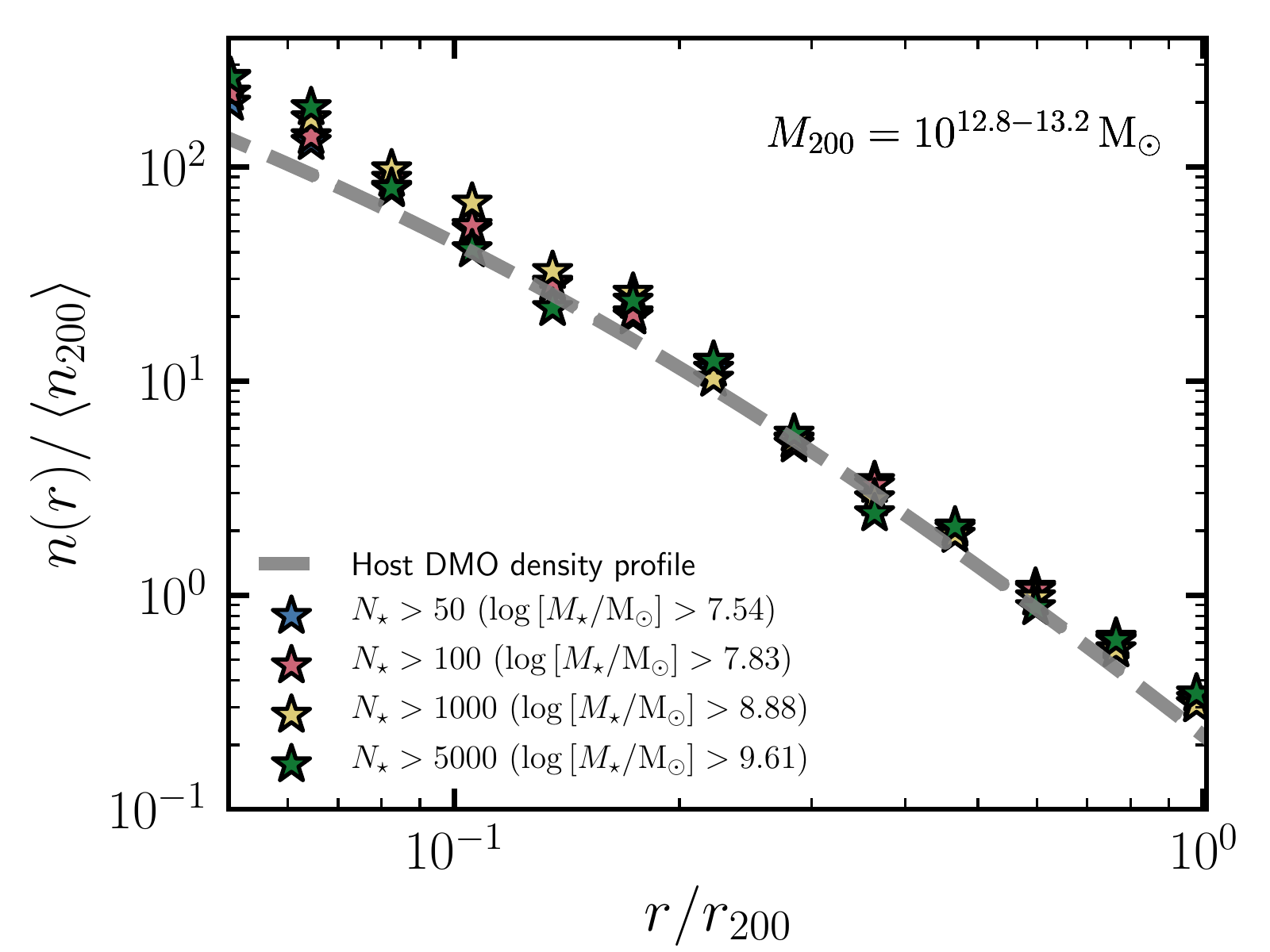}
    \caption{The radial number density profile of luminous satellites extracted from TNG100 for different choices of the minimum number of star particles, $N_\star$, in the galaxy at $z=0$. The red stars ($N_\star>100$) reproduce the result presented in the top right panel of Fig.~\ref{fig:sat_rad_distr}; the grey curve is identical to orange curve displayed in that figure.  Neither the shape nor the amplitude of the normalised profiles are affected strongly by the star particle threshold.}
    \label{fig:sat_nstarconv}
\end{figure}

Fig.~\ref{fig:sat_nstarconv} repeats the calculation presented in Fig.~\ref{fig:sat_rad_distr} for different choices for the minimum mass of galaxies used to construct the radial profile. For clarity, we show results solely for the host halo mass range $M_{200} = 10^{12.8-13.2}\,{\rm M}_\odot$, although our conclusions are unchanged for other host mass ranges. It is clear from this figure that the profiles for each mass cut is consistent with our fiducial choice of $N_\star>100$. Reassuringly, both the amplitude and shape of the {\it normalised} profiles is consistent across all satellite mass ranges, although there is a small deviation between individual profiles towards the centre of the host halo. Fig.~\ref{fig:sat_nstarconv} suggests that the similarity between the radial profile of galaxies and the host dark matter profile is not biased by the dominance of a particular mass range of satellite galaxies in the radial profile.

\bibliographystyle{mnras}
\bibliography{hod.bib}


\bsp	
\label{lastpage}
\end{document}